\documentclass[11pt]{article}

\usepackage[
  a4paper,
  left=2.8cm,
  right=2.8cm,
  top=2.5cm,
  bottom=2.5cm
]{geometry}

\usepackage[T1]{fontenc}
\usepackage{textcomp}
\usepackage{microtype}

\usepackage{graphicx}
\graphicspath{{figs/results/}}
\usepackage{booktabs}
\usepackage{longtable}
\usepackage{array}
\usepackage{tabularx}
\usepackage{ragged2e}
\usepackage{siunitx}
\usepackage{amsmath,amssymb}
\usepackage{caption}
\usepackage{subcaption}
\usepackage{setspace}

\usepackage{placeins}

\usepackage{algorithm}
\usepackage{algpseudocode}
\usepackage{listings}
\usepackage{verbatim}
\usepackage{tikz}
\usetikzlibrary{positioning, arrows.meta, shapes.geometric, fit, backgrounds}
\usepackage{forest}

\usepackage{authblk}

\usepackage[
  backend=biber,
  style=authoryear-comp,
  natbib=true,
  sortcites=false,
  maxcitenames=2,
  mincitenames=1,
  maxbibnames=99,
  giveninits=true,
  uniquename=init,
  uniquelist=false,
  date=year,
  urldate=iso,
  seconds=true,
  isbn=false
]{biblatex}
\AtEveryBibitem{%
  \iffieldundef{doi}{}{%
    \clearfield{urlyear}\clearfield{urlmonth}\clearfield{urlday}%
  }%
  \ifentrytype{article}{%
    \iffieldundef{doi}{}{\clearfield{eprint}}%
  }{}%
  \iffieldundef{doi}{}{\clearfield{url}}%
}

\usepackage{xurl}
\usepackage[hidelinks]{hyperref}
\usepackage[nameinlink,noabbrev]{cleveref}
\usepackage{orcidlink}

\Crefname{figure}{Fig.}{Figs.}
\Crefname{table}{Table}{Tables}
\Crefname{section}{Section}{Sections}

\title{3Dtrees.earth: community research infrastructure for tree and forest analysis using close-range LiDAR data}

\author[1,*,c]{Janusch Vajna-Jehle\,\orcidlink{0009-0008-9195-5633}}
\author[1,*]{Kilian Gerberding\,\orcidlink{0009-0002-5001-2571}}
\author[1]{Daniel Lusk\,\orcidlink{0009-0002-9745-5011}}
\author[1]{Teja Kattenborn\,\orcidlink{0000-0001-7381-3828}}
\author[1]{Julia Sophie Gäßler\,\orcidlink{0009-0002-6393-8218}}
\author[1]{Anna Göritz\,\orcidlink{0000-0001-9844-0896}}
\author[1]{Luis Kremer\,\orcidlink{0009-0007-1613-5793}}
\author[1]{Charlotte Lange\,\orcidlink{0009-0000-9731-525X}}
\author[1]{Aaron Sheppard}
\author[1]{Salim Soltani\,\orcidlink{0000-0003-2797-4131}}
% ...

\affil[1]{Chair of Sensor-based Geoinformatics (geosense), University of Freiburg, Freiburg im Breisgau, Germany}
% ...
\affil[*]{These authors contributed equally to this work.}
\affil[c]{Corresponding author: Janusch Vajna-Jehle, \href{mailto:janusch.jehle@geosense.uni-freiburg.de}{janusch.jehle@geosense.uni-freiburg.de}.}

\begin{document}

\date{}
\maketitle
\clearpage
\begin{abstract}

From individual trees to forest stands, three-dimensional structure shapes carbon, water, and energy flows, habitat conditions, and resource availability. Yet tree and forest structural attributes remain difficult to assess consistently across space and time. Close-range LiDAR (light detection and ranging) records detailed 3D structure from branches and trees to forest plots, including trees outside forests. However, broader reuse is constrained when data, acquisition context, processing, computation, and access conditions are maintained separately. We present 3Dtrees.earth, an operational community research infrastructure that brings close-range LiDAR datasets, analysis tools, and computing resources together while linking original observations to their metadata and processing outputs. As of 23 October 2026, the platform comprised 2{,}760 datasets, including 2{,}481 publicly visible datasets. These point cloud datasets span 46 countries on all six inhabited continents and include terrestrial, mobile, and uncrewed aerial laser scanning. Users can contribute and discover datasets, run standardized processing workflows on shared computing infrastructure, inspect point clouds and model predictions in the browser, and access data and processing outputs together with citation information. Workflows combine rule-based and machine-learning methods, with tools updated as methods advance. Outputs include raster products, individual-tree and semantic segmentation, and tree- and stand-level metrics. Original point clouds and processing records are retained, enabling reprocessing with improved tools while preserving provenance. With broader coverage, curation, and validation, this shared resource could extend individual surveys into a basis for model evaluation and benchmarking, cross-site synthesis, calibration and validation of larger-scale Earth observation products, and macroecological analysis.

\end{abstract}

\section{Introduction}
\subsection{Close-range LiDAR reveals tree and forest structure}
Trees form complex three-dimensional structures in forests, savannas, agricultural landscapes, and urban areas.
Forests alone cover roughly a third of the Earth's land surface, regulate global carbon, water, and
energy cycles, and sustain a large share of terrestrial biodiversity
\citep{kriegerEconomicValueForest2001,groombridgeWorldAtlasBiodiversity2002a,pan2013structure,faoGlobalForestResourcesAssessment2020,santoroGlobalForestAboveground2021}.
Ecosystem processes in tree-dominated ecosystems are inherently three-dimensional, with structural complexity shaping habitat availability, microclimate, biodiversity, ecosystem fluxes, and resilience to climate change \citep{ehbrecht2021global,linesShapeTreesReimagining2022,atkins2023integrating,schnabelTreeDiversityTemperature2025}. At the same time, tree mortality, forest degradation, and forest management are altering ecosystem structure \citep{ipbesGlobalAssessmentReport2019, mcdowellPervasiveShiftsForest2020, schieferLargescaleRemoteSensing2025}. Furthermore, climate-driven disturbance regimes are intensifying \citep{seidlForestDisturbancesClimate2017, mahechaBiodiversityClimateExtremes2024}, creating an urgent need to quantify dynamics in tree and forest structure consistently across large areas \citep{networkGlobalUnderstandingTree2025}. However, obtaining comparable structural observations across sites and through time remains challenging \citep{fassnachtRemoteSensingForestry2024}.

Traditional field methods capture that geometry only to a limited extent.
Forest inventories provide reference measurements for many forest attributes, but collecting them is labor-intensive, limiting the number of sites that can be surveyed and how frequently measurements can be repeated
\citep{tomppoCombiningNationalForest2008, averyForestMeasurementsFifth2015}. National forest inventories
use country-specific sampling designs, with differing plot sizes, sampling densities, and revisit intervals \citep{vidalNationalForestInventories2016}. These designs support inference at broad reporting scales but provide limited resolution in terms of fine-grained spatial variation and short-term structural change
\citep{kandlerDesignSecondGerman2009, neshaAssessmentDataSources2021}. Access to exact national forest inventory plot locations is often restricted, which can complicate integration with remote sensing observations
\citep{gesslerFindingBalanceOpen2024,schumacherFederatedLearning2026}. Plot locations are also often poorly georeferenced because dense canopies weaken satellite-positioning signals \citep{maackModellingStandingTimber2016}. Field assessments that rely on expert judgment rather than direct measurement can also be subject to observer bias \citep{innesForestHealthObserver1988,bacaroObserverExpertiseBias2026}. Therefore, field-based monitoring is limited by the range of structural metrics it
can quantify and the sampling error caused by restricted spatial and temporal coverage
\citep{mcdowellPervasiveShiftsForest2020, networkGlobalUnderstandingTree2025}.

Close-range LiDAR (light detection and ranging) fundamentally changes this measurement paradigm: rather than inferring tree structure from a limited set of plot-level measurements, it resolves tree geometry directly in three dimensions. Terrestrial laser scanning (TLS), mobile laser scanning (MLS), and uncrewed aerial laser scanning (ULS) systems detect reflected laser pulses and record up to millions of three-dimensional point coordinates per second. The resulting point clouds reconstruct the explicit three-dimensional geometry of trees and forest stands \citep{liangTerrestrialLaserScanning2016,caldersTerrestrialLaserScanning2020,liangCloseRangeRemoteSensing2022,coopsAdvancesLaserScanning2025}, and repeated acquisitions can document structural change over
time \citep{sheppardTerrestrialLaserTreeGrowth2017,gassilloudTrackingForestDensity2026}. Three-dimensional observations enable detailed estimates of tree dimensions and crown structure \citep{owenCommonFieldMeasures2024}, as well as foliage density and biomass \citep{caldersNondestructiveEstimatesAboveground2015,belandMappingForestLeaf2021}. Close-range LiDAR therefore provides structural variables that are difficult to obtain from conventional surveys \citep{caldersNondestructiveEstimatesAboveground2015,belandMappingForestLeaf2021,pulitiBranchPoseNetCharacterizingTree2024,feiglCloseRangeRemoteSensing2025}. These observations complement established forest inventories rather than replace their long-term measurements and sampling frameworks \citep{fassnachtRemoteSensingForestry2024}.

\begin{figure}[H]
  \centering
  \includegraphics[width=\textwidth]{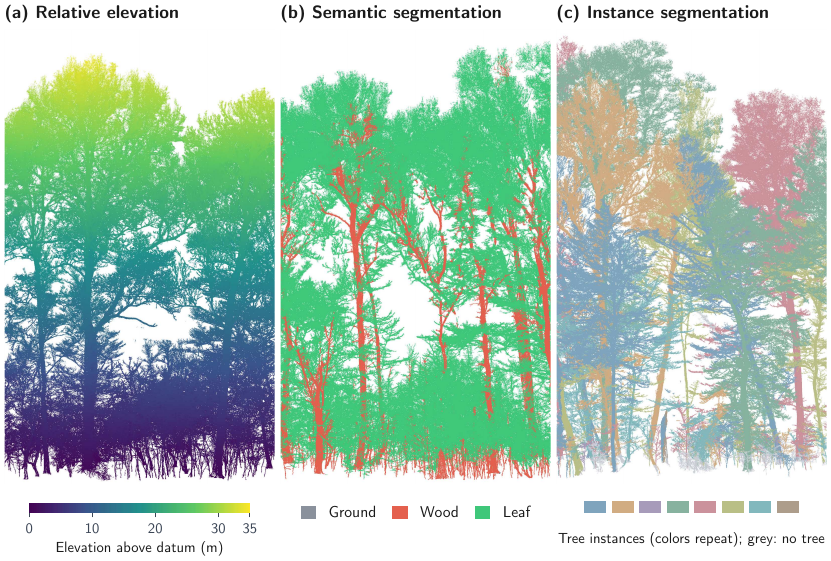}
  \caption[Data-derived point cloud and segmentation example]{Terrestrial laser scanning example (3Dtrees.earth dataset 3302, FOR-instance v2: \textit{BlueCat} training split\footnotemark): adjacent sections of a 66~m-long, 20~m-deep oblique transect, (a)~colored by height, (b)~colored by \textit{ForestMamba} \citep{nguyenForestMambaSparse2026} semantic predictions (ground, wood and leaf) and (c)~colored by \textit{RayCloudTools} \citep{devereuxRayExtractFastScalable2026} tree-instance predictions. Instance colors repeat, and gray points are not assigned to a tree.}
  \label{fig:point_cloud_segmentation_dataset_1367}
\end{figure}

A growing set of methods offers opportunities to translate LiDAR data into quantitative, tree- and stand-level assessments of forest structure and function, as illustrated by a dataset in \autoref{fig:point_cloud_segmentation_dataset_1367}. Separation of leaf and wood points supports estimates of leaf distribution and leaf area index \citep{brownLeafWoodClassification2025}, while instance segmentation of individual trees \citep{wielgoszSegmentAnyTreeSensorPlatform2024,xiangForestFormer3DUnifiedFramework2025}
supports forest inventories and the estimation of structural attributes \citep{pulitiFORinstanceUAVLaser2023, xiang2023accurateinstancesegmentationlargescale, xiangAutomatedForestInventory2024}. On delineated instances, species recognition and biomass estimation support biodiversity monitoring and carbon accounting \citep{caldersNondestructiveEstimatesAboveground2015,schindlerAllometricWildCherry2023,pulitiBenchmarkingTreeSpecies2025,coopsAdvancesLaserScanning2025,networkGlobalUnderstandingTree2025}. Reconstructed 3D structure informs radiative-transfer simulations \citep{frey3DReconstructionWood2025} and the study of competition for light \citep{kroner2026disentangling}. It also supports microclimate modeling \citep{schnabelTreeDiversityTemperature2025} and the estimation of finer-grained attributes such as tree age \citep{pulitiFORageBenchmarking2026} and habitat quality \citep{holland2024terrestrial}. Within surveyed areas, these methods can enable highly detailed, spatially explicit
characterization of individual trees and stands, thereby extending the range of structural attributes
accessible through conventional field inventories.%

These advances highlight the potential of close-range LiDAR, but their broader application depends on understanding and improving methodological transferability across biomes, forest types, and acquisition settings. This requires diverse training and evaluation datasets, making the fragmentation of existing data a practical barrier to further progress \citep{CHERLET2026230}.

\footnotetext{\url{https://3dtrees.earth/datasets/3302}}
\subsection{Fragmentation of data and tools limits reuse}
Individual LiDAR surveys reveal forest structure in detail, but comparisons across sites and time require a broader set of observations. Pooling datasets across sensor types, forest types, regions, and time can support cross-site synthesis, method development, and links to larger-scale Earth observation (EO) products.

Implementing research with aggregate LiDAR data requires this data to be findable, accessible, interoperable, and reusable (FAIR) \citep{wilkinson2016fair}, which requires communication of the acquisition, processing, and quality of the dataset \citep{schiefer2023uav,fassnachtRemoteSensingForestry2024,maedaExpandingForestResearch2025}. Currently, the path from individual LiDAR datasets to scientific reuse runs through disconnected systems: data, documentation, processing tools, computational resources, and
results are commonly maintained separately. Consequently, researchers must find and move data between
environments, reconstruct how they were produced, and assemble the tools needed for analysis.
This fragmentation narrows the data available for developing transferable methods and synthesis studies \citep{kattenbornReviewConvolutionalNeural2021}.

Existing initiatives and resources cover different parts of this path. General-purpose archives such as Zenodo or PANGAEA preserve and publish research data \citep{zenodo2013, feldenPANGAEADataPublisher2023}; tools such as
\textit{PC2RCHIVE} help researchers prepare point clouds and associated metadata for archiving \citep{farkasPC2RCHIVE2024}. SmartForest and ScanForest make shared code and forest point cloud datasets discoverable \citep{smartforestCodesDatasets2026, scanforestData2026}; Global TLS documents plot-scale terrestrial laser scanning activity and indicates where data are available\footnote{\url{https://www.global-tls.net}}, while OpenTopography provides mature infrastructure for airborne laser scanning (ALS) point cloud discovery, access, and processing \citep{crosbyOpenTopography2020}.

Task-specific labeled datasets support method development and evaluation: FOR-instance provides a
UAV laser scanning benchmark for semantic and instance segmentation, FOR-species20K supports
training and benchmarking of tree-species classifiers from single-tree point clouds, and FOR-age
benchmarks individual-tree age estimation from dense laser scans
\citep{pulitiFORinstanceUAVLaser2023, pulitiBenchmarkingTreeSpecies2025,
pulitiFORageBenchmarking2026}. SegmentedForests adds manually labeled ground-based point clouds for semantic segmentation of forest scenes \citep{lainoSegmentedForestsLabelledDataset2025}. TreeScanPL10k provides annotated terrestrial laser scanning point clouds for individual-tree segmentation and species classification \citep{sterenczak_2026_19127709}. BorFIT contributes manually segmented individual trees from northern boreal forests for developing and evaluating tree-segmentation methods \citep{schladebachBorFIT2025}.

Combining these datasets and reproducing analyses still requires researchers to connect point cloud files with acquisition metadata and processing workflows. Point clouds can be distributed in various open or proprietary file formats, and information regarding how datasets were acquired, registered, georeferenced, filtered, and processed is often implicit or distributed across file metadata, README documents, and separate publications \citep{liangCloseRangeRemoteSensing2022, ozturkBarriersGaps2026}. For close-range LiDAR, differences in scan positions, occlusion, and point density affect how completely tree structures are captured and which structural attributes can be reliably derived \citep{caldersTerrestrialLaserScanning2020, gassilloudOcclusionMapping2025, kukenbrinkInsightsUnseen2025}. 

Reproducing the processing applied to those data poses a further barrier. Highlighting such challenges, a review within the COST Action 3DforEcoTech cataloged 24 software solutions for processing
ground-based forest point clouds \citep{murtiyosoReviewSoftwareSolutions2024}.

While this diversity reflects an active research community, differences in software environments, input requirements, dependencies, and documentation can make it difficult to combine tools, trace processing steps, and reproduce analyses. Using and combining these tools often requires specialist expertise. Furthermore, computational demands restrict access to many recent
approaches \citep{ozturkBarriersGaps2026}. Projects such as \textit{ForestSens} complement these efforts by providing cloud access to forest point cloud algorithms \citep{smartforestForestSens2026}, but running tools in the cloud alone does not connect processing runs to the broader dataset metadata and provenance needed for interpretation and reuse.

Observation networks and research platforms provide useful examples of how different parts of this process can be connected through shared infrastructure. The pioneering PhenoCam Network uses common protocols and openly shared repeat photography to support phenological research across sites; its development also illustrates the need for sustained collaboration and funding \citep{richardsonPhenoCamEvolving2023}. OpenML links datasets to tasks, implementations, and recorded results,
making it easier to compare methods on common data \citep{bischlOpenMLInsights2025}. iNaturalist
connects plant and animal species observations to community identification and revision, allowing contributions to be improved through use \citep{campbellIdentifyingIdentifiers2023}. The Global Biodiversity Information Facility (GBIF) indexes species records from many distributed providers through common infrastructure for discovery, access, citation, and attribution \citep{robertsonGBIFIntegrated2014}.\footnote{\url{https://www.gbif.org/}} The platform deadtrees.earth\footnote{\url{https://deadtrees.earth/}} provides data publication and sharing of close-range RGB data from UAVs as well as inference on forest cover and dead wood mapping \citep{mosigDeadtreesearthOpenaccessInteractive2026}. These platforms serve different purposes, but each reduces the work required to progress from submitting or finding a contribution to processing, analyzing, or improving it. The remaining challenge is to establish these connections for close-range LiDAR data.

\subsection{3Dtrees.earth: Concept and goals}

We present 3Dtrees.earth, an operational research platform that unifies research data management, cloud-based processing, and web-based visualization of close-range LiDAR datasets, linking uploaded data, metadata, and derived products for provenance tracking and method comparison throughout.  The core data focus of 3Dtrees.earth is point clouds produced from terrestrial laser scanning (TLS), mobile laser scanning (MLS), and uncrewed aerial laser scanning (ULS) sensors, given their ability to resolve fine tree and forest structures inaccessible to larger-scale sensors. We begin by detailing the dataset life cycle in Section~\ref{sec:current_state}. We describe how datasets are uploaded, processed, inspected, and accessed through 3Dtrees.earth, followed by a summary of the current dataset collection. We also discuss how the platform is governed and a plan for sustainable maintenance and growth. Looking forward, Section~\ref{sec_outlook} examines the scientific opportunities and limitations of this growing collection, together with the curation, validation, and coverage needed to support its use and maximize its utility for researchers, developers, and practitioners alike.

\section{Platform implementation: From data and method contribution to reuse}
\label{sec:current_state}

\begin{figure*}
  \centering
  \includegraphics[height=0.76\textheight,keepaspectratio]{%
    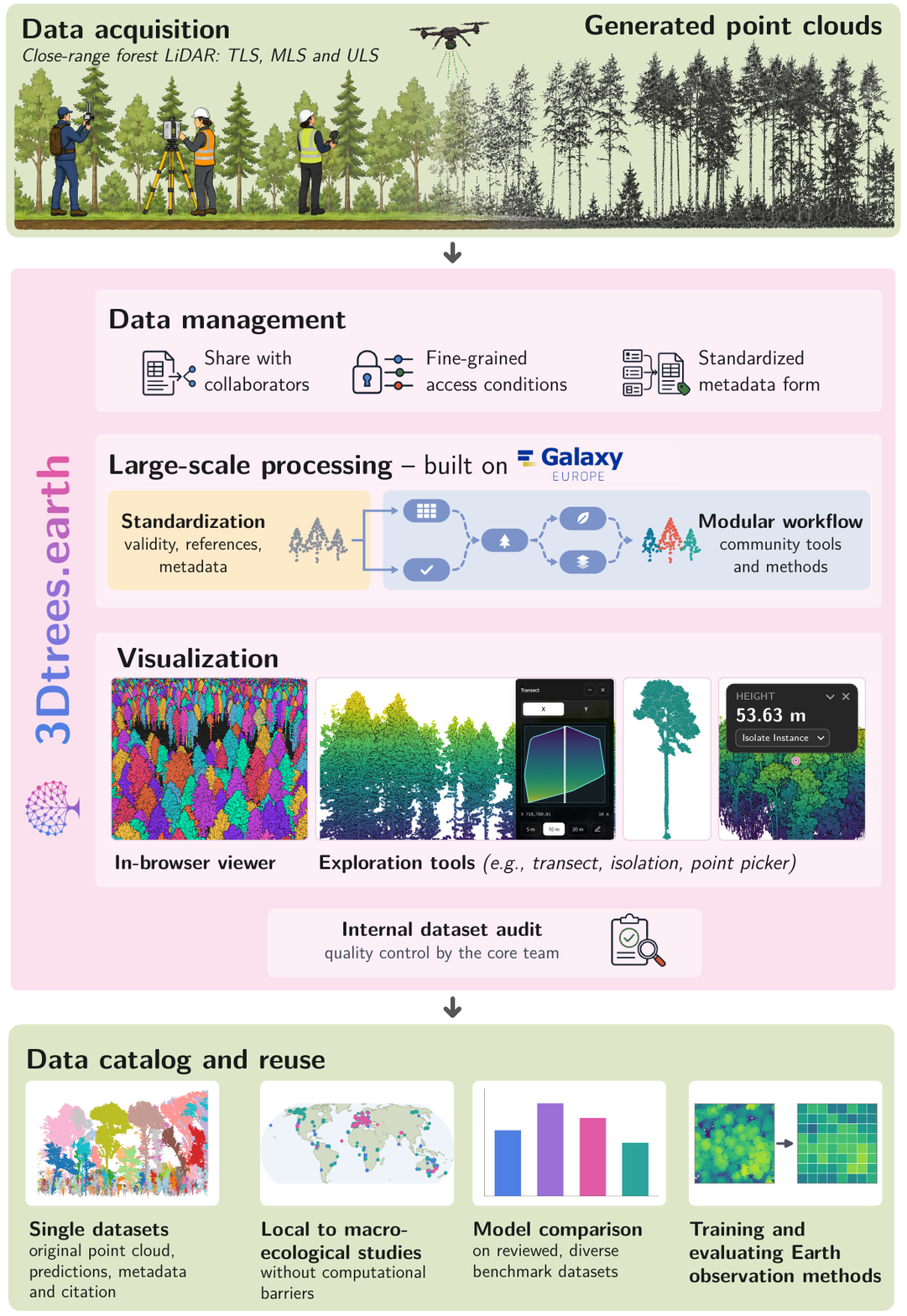}
  \caption[3Dtrees.earth: From close-range forest LiDAR to reuse]{Overview of
  3Dtrees.earth from close-range forest LiDAR acquisition through data
  management, Galaxy Europe processing, web-based inspection, and dataset
  access. Operational capabilities shown include structured metadata and
  access conditions, standardization, tree-instance prediction, conditional
  species prediction within the model's European scope, raster products,
  single-tree structural metrics, browser visualization,
  internal quality control, and prepared per-dataset archives. Dataset releases
  provide persistent DOIs. Broader curation, curated releases
  with reviewed labels, benchmark datasets, and wider
  Earth observation applications are planned. Selected schematic illustrations
  were created using OpenAI image-generation tools and edited by the authors.}
  \label{fig:3dtrees_main}
\end{figure*}

3Dtrees.earth is a web-based application that enables researchers to manage, analyze, and share close-range LiDAR point clouds. Standardized workflows provide access to tools for analyzing tree and forest structure on publicly funded cloud computing infrastructure, allowing researchers to obtain processed point clouds and derived products without having to set up or maintain their own processing environments.

Users can discover public datasets on an interactive map, inspect and compare point clouds and model predictions in the browser, and download data together with available processing records (\autoref{fig:3dtrees_main}). Each uploaded dataset retains its acquisition metadata, contributor information, and access settings. Derived products remain linked to the original point clouds and the processing records that describe how they were generated. Retaining the original data allows datasets to be reprocessed as methods improve or new applications emerge. The main software components and their connections are shown in Appendix~\ref{app:platform_architecture} (\autoref{fig:platform_architecture}).

\subsection{Data contribution and discovery}
\label{sec:data_contribution_management}

After signing in, a user contributes close-range scans as one or more point cloud files in the open LAS or LAZ formats. Preprocessing to translate the raw scanning data into a point cloud must be done prior to upload, since this step usually requires specialized proprietary software provided by the sensors' manufacturer. Here, ``original point clouds'' refers to the preprocessed files uploaded by contributors before processing on 3Dtrees.earth.

3Dtrees.earth organizes data into datasets, which are discovered, processed, and accessed individually. A \textit{dataset} corresponds to a single LiDAR survey, covering one contiguous area with one acquisition date. A LiDAR survey may produce several files, for example, from different scanning positions or subareas, and these can be uploaded simultaneously to form one dataset sharing a single metadata record. Files from different scan positions must be co-registered before upload; 3Dtrees.earth does not currently perform this step. Contributors can additionally group related datasets, such as surveys from different sites or dates within one project, into a \textit{collection}.

A dataset's metadata record is built in two parts: 3Dtrees.earth extracts point cloud attributes, including the coordinate reference system, automatically, while the user supplies acquisition details such as sensor information and scanning pattern during upload. The record also holds authorship and funding information, and users can attach supplementary material such as related documentation. Two features reduce the effort of entering this information. First, a user can supply the digital object identifier (DOI) of an existing publication or repository record; the platform then retrieves the available descriptions, authors, Open Researcher and Contributor ID (ORCID) identifiers, and funding information from that source while preserving the link to the original work. Users can also have a DOI provisioned for their dataset (Section~\ref{sec:dataset_access_publication}). Second, reusable templates can be used to carry shared metadata across a series of related datasets, reducing repeated data entry and improving consistency between uploads. Appendix~\ref{app:dataset_metadata} lists the metadata fields and distinguishes required, optional, and automatically extracted information.

Access is governed on a per-dataset basis according to the principle ``\textit{as open as possible, as closed as necessary}'' \citep{europeancommissionGuidelinesFAIRData2016}. The uploading user selects one of three access levels: \textit{Public} datasets are openly discoverable, inspectable, and downloadable; \textit{View-only} datasets remain discoverable and inspectable, but downloads require explicit access; and \textit{Private} datasets are accessible only to authorized users.

Users can share datasets or collections with registered collaborators, granting permission to download data, edit metadata, or manage access. This enables collaboration without requiring public release to support constraints such as research status, inventory regulations, or commercial confidentiality. Additionally, a dataset's location can be provided as approximate or even hidden instead of as precise coordinates for sensitive sites. While visibility and access management govern who can obtain a dataset, users must also select a license, with CC-BY-4.0 recommended. Licenses are packaged with dataset downloads, governing dataset usage outside the platform.

\begin{figure*}[tbp]
  \centering
  \input{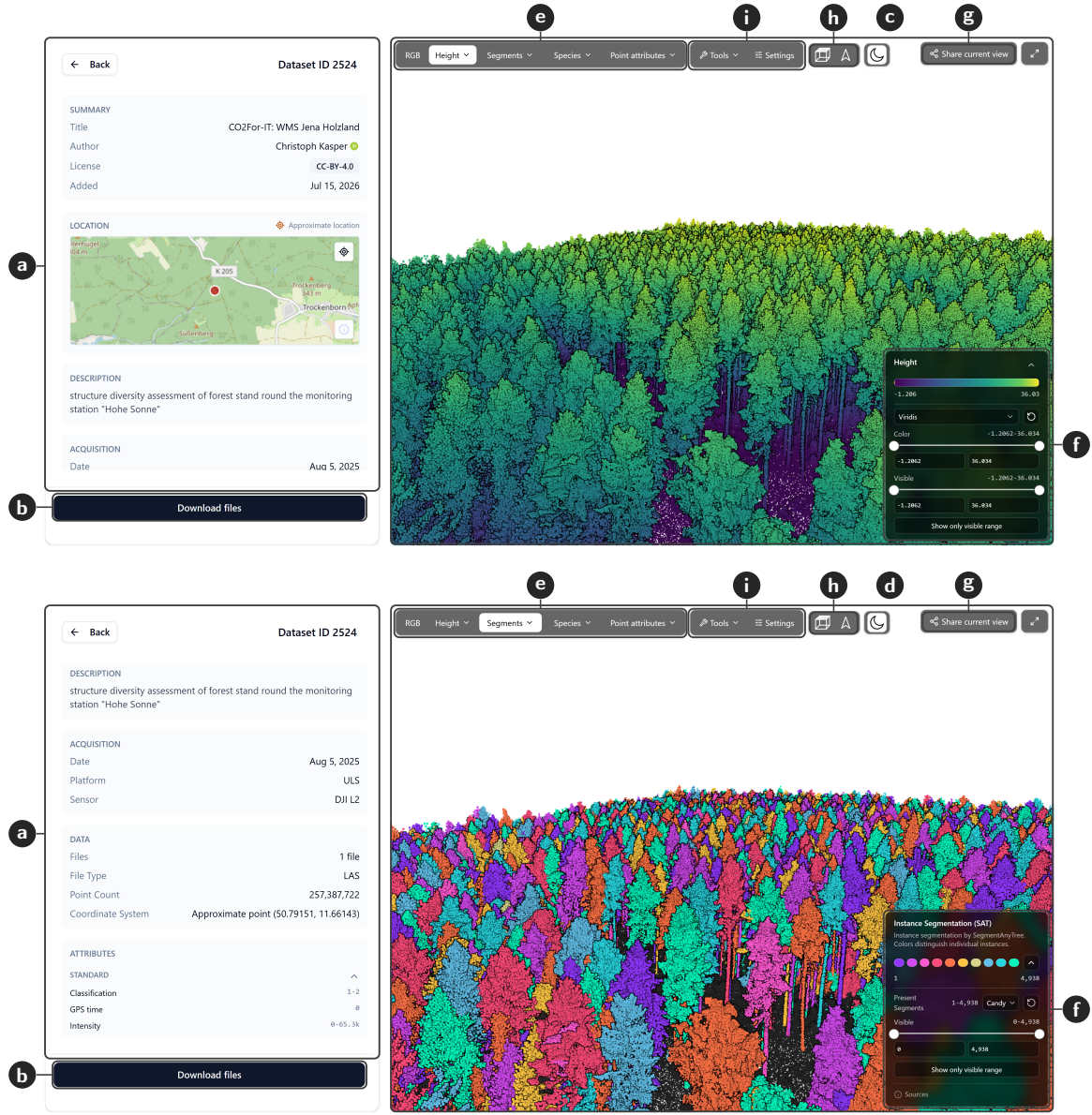}
  \caption[The 3Dtrees.earth dataset detail page]{The 3Dtrees.earth dataset
  detail page (3Dtrees.earth dataset 2524, CO2For-IT: WMS Jena Holzland\footnotemark) brings together a dataset's metadata (a), the
  dataset download (b), and web-based point cloud inspection,
  here colored by (c)~height and (d)~predicted tree instance. The toolbar
  (e) selects the point attribute and coloring; the expanded legend (f)
  shows the color scale with its color-range and visible-range controls.
  Any view, including camera position and coloring, can be shared as a link
  (g). Besides rotating, panning and zooming, the view can be switched
  between perspective and orthographic projection and turned to face north
  (h). The tools menu (i) offers distance measurement and transects, and
  the settings adjust point shape, size and the number of points loaded.
  In the lower row, the sidebar is scrolled to the acquisition and file
  information.}
  \label{fig:app_dataset_page}
\end{figure*}

Public and view-only datasets can be discovered via the dataset catalog. The catalog combines an interactive world map with a list of all datasets. Search and sorting tools help identify relevant datasets. Selecting a dataset opens its detail page (\autoref{fig:app_dataset_page}), which presents the metadata, any linked DOI, the location, and the available processing outputs alongside the web-based viewer, supporting the assessment of a dataset's suitability for a user's needs prior to download.

\footnotetext{\url{https://3dtrees.earth/datasets/2524}}
\subsection{Processing and provenance on Galaxy Europe}
\label{sec:galaxy_processing_outputs}

A common challenge for researchers analyzing large LiDAR datasets is access to sufficient computing resources, especially for demanding methods such as deep learning.
3Dtrees.earth therefore connects contributed point clouds with reproducible processing on Galaxy Europe\footnote{\url{https://galaxyproject.org/eu/}}, a research-computing infrastructure that federates European high-performance computing and cloud centers into shared computing capacity \citep{ thegalaxycommunityGalaxyAccessibleReproducible2026}. Running the most demanding stages on shared infrastructure removes the need for individual high-performance hardware. Modular workflows prepare datasets for analysis, derive analytical products, and generate assets for web-based inspection. The infrastructure supports an evolving combination of state-of-the-art methods rather than a fixed sequence of algorithms.

The analytical tools in 3Dtrees.earth currently include a range of conventional geometric and rule-based methods alongside machine-learning models. Because tools installed on the Galaxy servers are running in their own execution environments, workflows can incorporate methods implemented in different programming languages, including R, Python, C++, and Julia. This approach allows communities of method developers to freely develop new tools for segmentation, species classification, or structural analysis, which the 3Dtrees.earth team can integrate after technical review and testing. Datasets already held on the platform can then benefit from reprocessing as improved or more specialized methods become available.

\begin{figure}[p]
    \centering
    \centering
\input{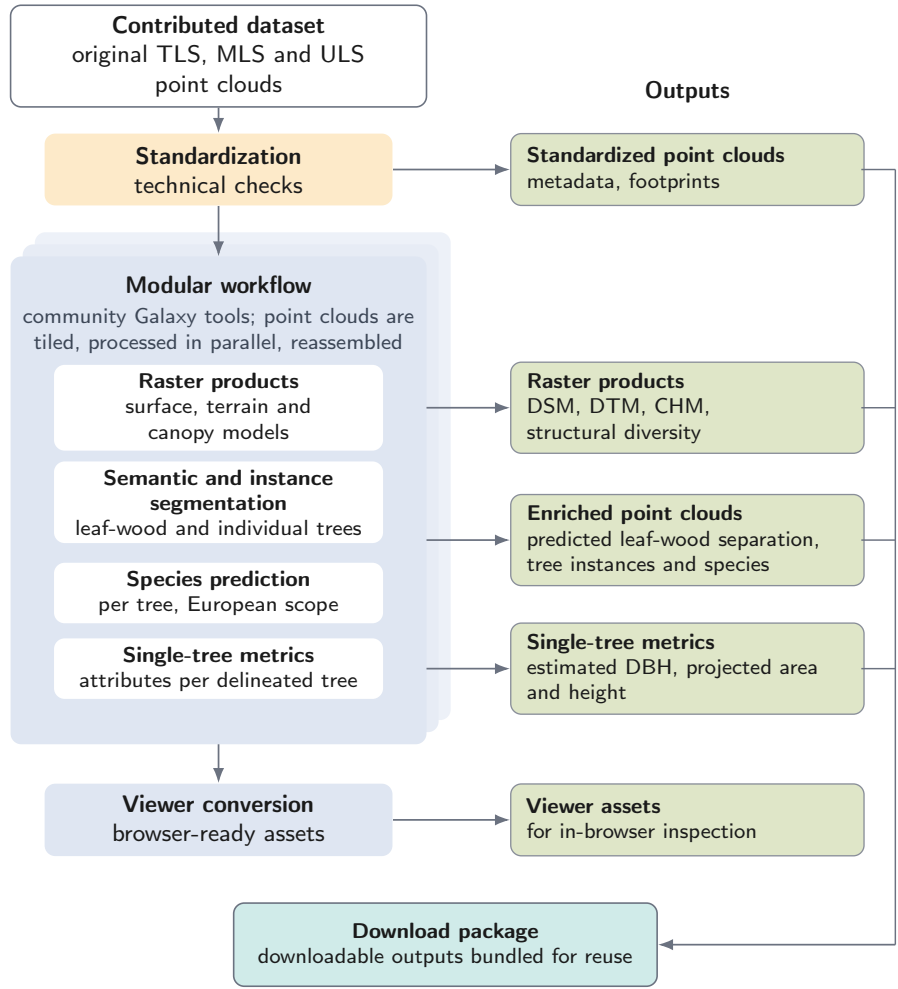}
\hyphenpenalty=10000
\exhyphenpenalty=10000
{%
\begin{tikzpicture}[
  x=1cm, y=1cm,
  font=\sffamily\footnotesize,
  io/.style={draw=ftLine, rounded corners=4pt, line width=0.6pt,
    fill=white, align=center, text width=5.5cm-12pt, minimum height=0.95cm,  %
    inner xsep=6pt, inner ysep=4pt, text=ftInk},
  proc/.style={draw=none, rounded corners=4pt,
    fill=ftWorkflow, align=center, text width=4.6cm-12pt, minimum height=0.95cm,  %
    inner xsep=6pt, inner ysep=4pt, text=ftInk},
  corestep/.style={draw=none, rounded corners=4pt, fill=white, align=center,
    text width=4.0cm, minimum height=0.8cm, inner xsep=5pt, inner ysep=3pt,
    font=\sffamily\scriptsize, text=ftInk},
  outchip/.style={draw=ftLine!80, rounded corners=4pt, line width=0.5pt,
    fill=ftField, align=left, minimum width=4.6cm, text width=4.25cm, minimum height=0.95cm,
    inner xsep=6pt, inner ysep=4pt, font=\sffamily\scriptsize, text=ftInk},
  pkgchip/.style={draw=ftLine, rounded corners=4pt, line width=0.6pt,
    fill=ftTeal!22, align=center, minimum width=5.6cm, minimum height=1.1cm,
    inner xsep=6pt, inner ysep=4pt, font=\sffamily\scriptsize, text=ftInk},
  arrow/.style={-{Latex[length=2.0mm]}, line width=0.65pt, draw=ftLine},
  buswire/.style={line width=0.55pt, draw=ftLine},
]

\node[io] (n0) at (0, 0.15) {\textbf{Contributed dataset}\\original TLS, MLS and ULS\\point clouds};
\node[proc, fill=ftStandard] (n1) at (0,-1.35) {\textbf{Standardization}\\technical checks};

\fill[rounded corners=4pt, ftWorkflow!55] (-2.43,-2.18) rectangle (3.07,-8.63);
\fill[rounded corners=4pt, ftWorkflow!75] (-2.59,-2.34) rectangle (2.91,-8.79);
\fill[rounded corners=4pt, ftWorkflow]    (-2.75,-2.50) rectangle (2.75,-8.95);

\node[font=\sffamily\footnotesize\bfseries, text=ftInk, anchor=north] at (0,-2.62) {Modular workflow};
\node[font=\sffamily\scriptsize, text=ftMuted, align=center, anchor=north] at (0,-3.05)
  {community Galaxy tools; point clouds are\\tiled, processed in parallel, reassembled};
\node[corestep] (cs1) at (0,-4.5) {\textbf{Raster products}\\surface, terrain and\\canopy models};
\node[corestep] (cs2) at (0,-5.75) {\textbf{Semantic and instance}\\\textbf{segmentation}\\leaf-wood and individual trees};
\node[corestep] (cs3) at (0,-6.95) {\textbf{Species prediction}\\per tree, European scope};
\node[corestep] (cs4) at (0,-7.95) {\textbf{Single-tree metrics}\\attributes per delineated tree};

\node[proc] (n4) at (0,-9.95) {\textbf{Viewer conversion}\\browser-ready assets};

\draw[arrow] (n0) -- (n1);
\draw[arrow] (n1) -- (0,-2.50);
\draw[arrow] (0,-8.95) -- (n4);

\node[font=\sffamily\footnotesize\bfseries, text=ftInk, anchor=south] at (6.2,-0.6) {Outputs};
\node[outchip] (o0) at (6.2,-1.35) {\textbf{Standardized point clouds}\\metadata, footprints};
\node[outchip] (o1) at (6.2,-4.5) {\textbf{Raster products}\\DSM, DTM, CHM,\\structural diversity};
\node[outchip] (o2) at (6.2,-6.25) {\textbf{Enriched point clouds}\\predicted leaf-wood separation,\\tree instances and species};
\node[outchip] (o3) at (6.2,-7.95) {\textbf{Single-tree metrics}\\estimated DBH, projected area\\and height};
\node[outchip] (o4) at (6.2,-9.95) {\textbf{Viewer assets}\\for in-browser inspection};

\draw[arrow] (n1.east)    -- (o0.west);
\draw[arrow] (2.75,-4.5) -- (o1.west);
\draw[arrow] (2.75,-6.25) -- (o2.west);
\draw[arrow] (2.75,-7.95) -- (o3.west);
\draw[arrow] (n4.east)    -- (o4.west);

\node[pkgchip] (pkg) at (3.0,-11.6) {\textbf{Download package}\\downloadable outputs bundled for reuse};
\foreach \y in {-1.35,-4.5,-6.25,-7.95} \draw[buswire] (8.5,\y) -- (8.95,\y);
\draw[buswire] (8.95,-1.35) -- (8.95,-11.6);
\draw[arrow] (8.95,-11.6) -- (pkg.east);
\end{tikzpicture}%
}%
    \caption[Current 3Dtrees.earth processing workflow]{The current default processing workflow on Galaxy Europe. An uploaded dataset is standardized and screened by automated technical checks, then tiled, processed in parallel, and reassembled by a modular workflow of community \textit{Galaxy} tools. Analytical steps produce digital surface models (DSM), digital terrain models (DTM), and canopy height models (CHM) as raster products, semantic and instance segmentation, per-tree species prediction within the model's supported species set and European scope, and single-tree metrics delivered as vector products; the segmentation and species results are model predictions rather than reviewed labels. Structural metrics are derived estimates whose reliability depends on the input data and processing methods. The illustrated tools and their combination may change as methods are added or updated. DBH denotes diameter at breast height. Viewer assets are prepared last. Each output is exported at the stage that produces it. Downloadable outputs are available in per-dataset download packages (Section~\ref{sec:dataset_access_publication}), while viewer assets are served for in-browser inspection only.}
    \label{fig:workflow}
\end{figure}

The current default workflow illustrates this approach (\autoref{fig:workflow}). Following upload, the dataset's point cloud files are submitted together, so that processing is scoped to the dataset rather than to individual files. \textit{Standardization} first reconciles LAS and LAZ conventions across the files, and automated technical checks then screen them before subsequent analysis (Section~\ref{sec:quality_assurance_audit}). Overview images support catalog browsing. \textit{SmartTile} divides the standardized point clouds into tiles, subsampling where needed, so that demanding processing steps can be run in parallel. Raster processing derives a series of raster products, including digital surface models, digital terrain models, canopy height models and a series of structural diversity metrics.

The following tools provide the current basis for tree-level analysis, which can expand with community contributions and advances in the state of the art: \textit{SegmentAnyTree} \citep{wielgoszSegmentAnyTreeSensorPlatform2024} and \textit{ForestMamba} \citep{nguyenForestMambaSparse2026} each predict individual tree instances, and each also returns semantic classes: tree and non-tree for \textit{SegmentAnyTree}; leaf, wood, and ground for \textit{ForestMamba}. \textit{RayCloudTools} \citep{loweRayCloudToolsConciseInterface2021} provides geometric processing methods, including \textit{RayExtract} for tree segmentation and reconstruction of woody structure \citep{devereuxRayExtractFastScalable2026}. \textit{DetailView} utilizes the resulting tree instances and adds species predictions within its supported species set and European scope \citep{pulitiBenchmarkingTreeSpecies2025}. \textit{CspStandSegmentation} \citep{freyJulFreyCspStandSegmentationSingle2024} derives structural estimates from delineated trees, including height, diameter at breast height, and horizontal projected area. The reliability of these estimates depends on input quality, segmentation, and the assumptions and parameter settings of the respective method.

The processed point cloud tiles are then merged into a Cloud Optimized Point Cloud (COPC). The merged point cloud retains the available semantic, tree-instance, and species predictions, together with prediction scores where available. \textit{PotreeConverter} \citep{schutzFastOutofCoreOctree2020} uses this point cloud to generate files optimized for browser-based visualization.

The point cloud products and visualization files are attached to the dataset alongside raster products, vector records of single-tree attributes, and overview images. The downloadable products are bundled into a download package (Section~\ref{sec:dataset_access_publication}), while the visualization files are used only by the web viewer.

Users publishing research based on 3Dtrees.earth need traceable processing records to report how results were generated and which tools and settings were used. \textit{Galaxy} automatically records this information for each processing step, including the tool and its version, the parameters supplied, the inputs and outputs, and the container environment in which the step ran. These records keep results traceable to the runs that produced them, including when datasets are reprocessed using different methods or settings.
A versioned workflow description\footnote{\url{https://3dtrees.earth/docs/methods/workflows/default}} documents the processing stages, tool settings, conditional steps, and expected outputs. Available dataset-specific processing records document the workflow, tool versions, and parameters used to generate the outputs and are included in the download package. Export as a complete machine-readable research object, such as Research Object Crate (RO-Crate), is planned \citep{soiland-reyesPackagingResearchArtefacts2022}.

\subsection{Web-based visualization}
\label{sec:inspection_quality_curation}

Scan geometry, structural coverage, and data quality are difficult to judge from metadata alone. 3Dtrees.earth therefore provides interactive, browser-based visualization of point clouds, available model predictions, and other tool outputs, allowing researchers to assess datasets without downloading the complete point clouds or installing specialist desktop software such as \textit{CloudCompare} \citep{CloudCompare}. This inspection helps identify issues in scanning or processing, such as occlusion, spatial artifacts, and implausible predictions. Comparing outputs from different methods can also reveal differences that require closer quantitative evaluation.

\begin{figure*}[tbp]
  \centering
  \input{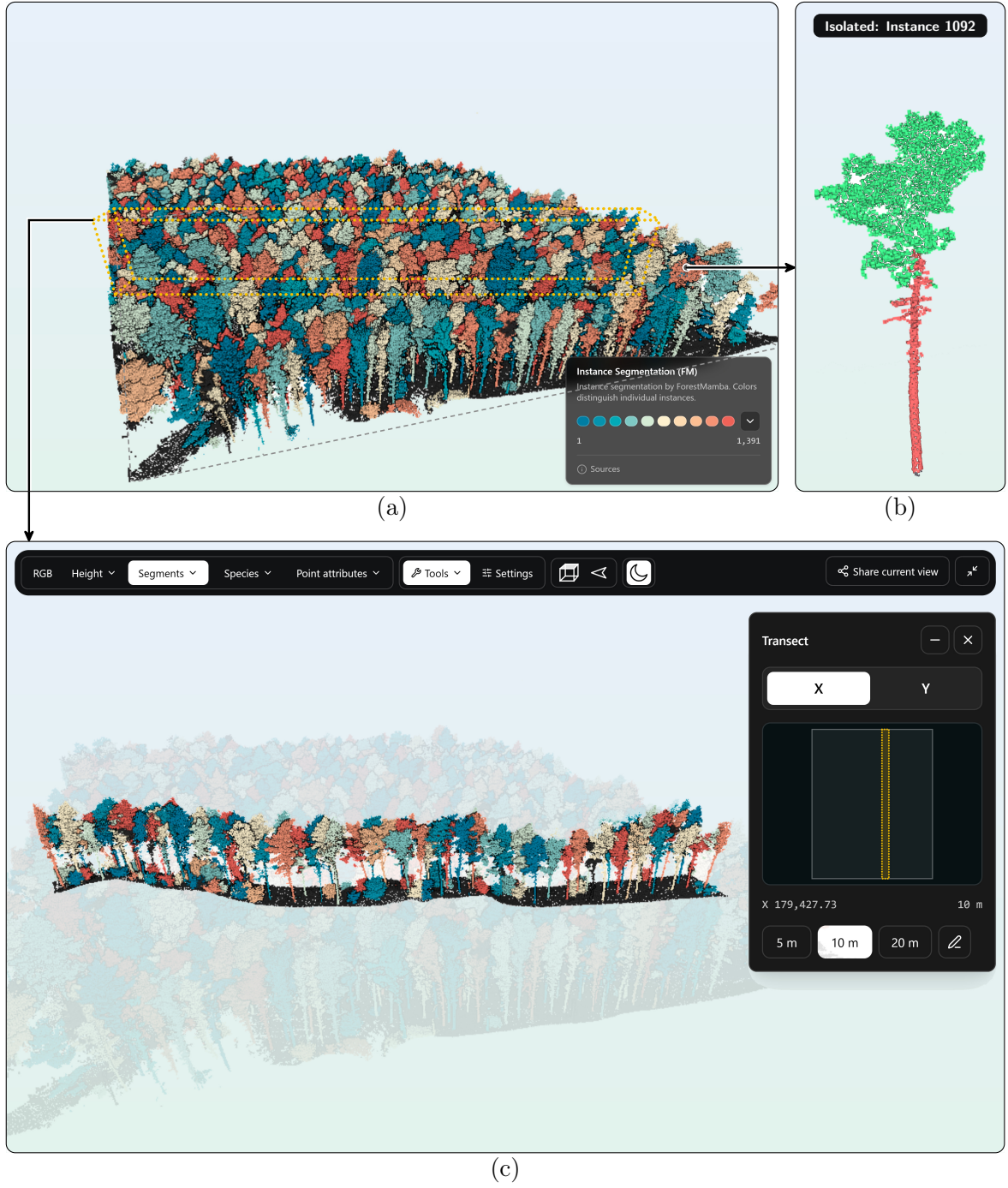}
  \caption[Browser-based inspection of point clouds and model predictions]{
  Browser-based inspection of a processed forest point cloud in the
  3Dtrees.earth viewer (3Dtrees.earth dataset 3261, Loobos fluxtower area scan\footnotemark), shown with its toolbar, color
  legend, instance-isolation bar and transect tool. (a)~The whole plot
  colored by predicted tree instance, with neighboring instances in
  contrasting colors. Arrows connect the marked transect to panel (c) and one predicted tree to panel (b).
  Panel (b) shows that tree isolated from the stand, colored by semantic prediction, with leaf points in green and wood points in red.
  (c)~The transect, of adjustable width, seen from the same viewpoint as (a)
  over the faded plot, exposes vertical stand structure and the spatial
  arrangement of predicted instances. Together, these views support the assessment of structural
  coverage, occlusion, and prediction plausibility without specialist local
  software.}
  \label{fig:app_transect}
\end{figure*}

The \textit{Potree}-based viewer \citep{Potree} streams the visualization files prepared during processing (Section~\ref{sec:galaxy_processing_outputs}; \autoref{fig:app_dataset_page}). It supports rotating, zooming, and panning through the three-dimensional scene. Point clouds can be colored and filtered by RGB where available, by height, by instance and species predictions, and by other point attributes present in the original point cloud or returned by the processing workflows. Neighboring predicted instances are assigned contrasting colors to keep them visually distinct (\autoref{fig:app_transect}). Adjustable transects expose vertical structure, canopy layering, and understory. Individual trees can be isolated to inspect their properties, including species predictions and, where available, prediction scores. Any view can be captured as a shareable link.

Visual inspection also provides the basis for structured dataset audits (Section~\ref{sec:quality_assurance_audit}).
\footnotetext{\url{https://3dtrees.earth/datasets/3261}}
\subsection{Quality assurance and dataset audit}
\label{sec:quality_assurance_audit}

Assessing a dataset's suitability for reuse requires both technical checks and examination of the point clouds and available predictions. The automated checks that run during processing record coordinate-reference information, spatial extent, point counts, overlap between files, and point-attribute consistency. A structured dataset audit complements these checks by documenting spatial coverage, acquisition context, and limitations relevant to interpretation and reuse.

Reviewers from the 3Dtrees.earth team examine datasets in a specialized audit view of the web viewer. They describe the dataset's acquisition and scene context, delineate usable areas and exclusions, and record occlusion, disturbance, metadata limitations, and descriptive tags. They also assess the plausibility of the available tree-instance and species predictions. These observations describe potential uses and limitations rather than assigning a single quality score. Quantitative assessment of prediction accuracy requires comparison with measurements or a defined reference standard. A second reviewer accepts a completed audit or returns it for revision, and any metadata corrections made during review remain visible in the dataset history. A common protocol and second review support consistency across assessments.

While audit records are currently internal, planned developments will make the findings available on dataset pages, support catalog filtering and dataset discovery, and include quality assessment information with downloads. A planned reporting function will also allow users to flag processing artifacts, implausible predictions, or metadata errors. The core team will review these reports and update the audit record where warranted.

The audit protocol is being developed with community partners and will evolve as the catalog grows and reuse experience accumulates. Because manual review requires time and expertise, we are exploring ways to automate parts of the assessment while maintaining reliable data-quality judgments. The current audit workflow and its status are documented on the platform\footnote{\url{https://3dtrees.earth/docs/methods/quality-audit}}.

\FloatBarrier
\subsection{Dataset access and published datasets}
\label{sec:dataset_access_publication}

Authenticated users can download any public dataset, as well as view-only or private datasets for which they have read access. They can select the original point clouds, available prediction-enriched point clouds, or both, along with supplementary material where available. Download packages always include the dataset's selected license, a README, a BibTeX citation, DataCite-formatted citation metadata, and available processing records in human- and machine-readable form. Files are accessed through temporary download links and can be saved as ZIP files directly from the browser or by using resumable download commands.

3Dtrees.earth enables contributors with administrative permission for the selected datasets to create citable dataset releases through FreiData, a research-data repository provided by the University of Freiburg\footnote{\url{https://freidata.uni-freiburg.de}}. A dataset release can contain an individual dataset, a collection, or a selection of datasets, together with their original point cloud files, metadata, and citation information. Contributors confirm their authority to publish the data and submit a draft for review by the core team. Acceptance triggers DOI registration and publication to FreiData, providing a persistent reference for citing and reusing the released data by the community.

\FloatBarrier
\subsection{Current catalog holdings and use}
\label{sec:platform_scope}

3Dtrees.earth opened to contributors in late 2025 and entered routine operation in early 2026. Within that first year, LiDAR datasets from many independent projects accumulated into a shared resource.

\begin{figure}[!tbp]
  \centering
  \includegraphics[width=\linewidth]{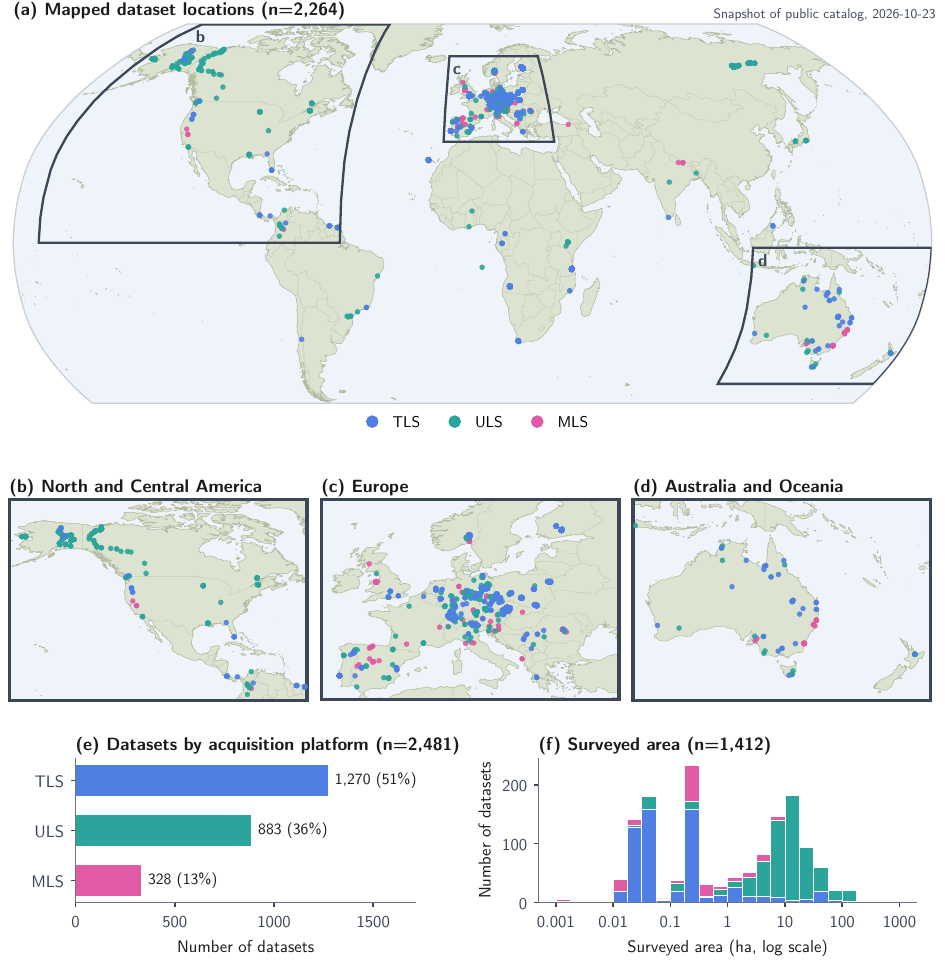}
\caption[Current holdings of 3Dtrees.earth]{Current holdings of 3Dtrees.earth as
of 23 October 2026. Panel (a) maps 2{,}264 dataset locations;
boxes mark the regions enlarged in (b) North and Central America, (c) Europe,
and (d) Australia and Oceania. Locations are centroids of the dataset footprint (1{,}412
datasets) or approximate coordinates where no footprint is available or the
contributor has chosen to hide the exact site (852 datasets). For the remaining 217 datasets no location information is available. Panel (e) shows
the acquisition platform of the 2{,}481 visible catalog datasets: 1{,}270
terrestrial (TLS), 883 uncrewed aerial (ULS), and 328 mobile laser scanning
(MLS) datasets. Panel (f) shows the distribution of surveyed area for the
1{,}412 datasets with a usable footprint, on a logarithmic hectare scale and stacked
by acquisition platform.}
  \label{fig:platform_current_holdings}
\end{figure}

By 23 October 2026, the catalog comprised 2{,}760 datasets, of which 2{,}481 were publicly visible (2{,}093 public and 388 view-only) and 279 private, involving 755 listed authors. Recorded point counts for 2{,}395 visible datasets total approximately 413.8~billion, and recorded file sizes for 2{,}364 visible datasets sum to approximately 5.32~TB. The 1{,}412 visible datasets with a usable footprint have a summed footprint area of 14{,}858~ha. Visible datasets span all three close-range acquisition modalities, comprising 1{,}270~TLS, 883~ULS, and 328~MLS datasets across sites, regions, and forest types (\autoref{fig:platform_current_holdings}).

Archive requests provide a further measure of operational activity. By 23 October 2026, 161 users had submitted 3{,}300 archive requests covering 1{,}679 datasets.

These figures describe the catalog at the reporting date; current statistics are available on the platform\footnote{\url{https://3dtrees.earth/stats}}.

The mapped holdings extend across 46 countries on all six inhabited continents. Of the 2{,}257 mapped records assigned to a country, approximately 73.6\% are in Europe, 11.7\% in North America, and 9.2\% in Africa, with smaller holdings in Oceania, Asia, and South America.

Several substantial collections illustrate the range of data available. TreeScanPL10k contributes 271 annotated TLS datasets from Central Europe \citep{sterenczak_2026_19127709}, while 171 TLS datasets capture woodland sites in Angola and Tanzania \citep{godleeSouthernAfricanWoodland2021}. The Alfred Wegener Institute's ULS collections cover northern boreal and tundra sites in Yakutia, Alaska, and northwestern Canada \citep{kruseYakutiaLidar2023,kruseEasternAlaskaLidar2025,kruseWesternAlaskaLidar2025,kruseCanadaLidar2025}. PathFinder superplots add MLS observations from European forest sites \citep{pathfinderSuperplots2026}. TomoSense combines TLS, MLS, and ULS observations within one temperate-forest collection, and LiPheStream contributes an 18-month point cloud time series from boreal Finland \citep{tebaldiniTomoSenseUnique3D2023,wittkeLiPheStream18monthHigh2024}. Together, these collections bring complementary observations together for comparisons across acquisition platforms, forest types, regions, and time.

The catalog's coverage reflects where contributing research groups work, which sensors they use, and which field sites they can access, rather than systematic sampling across regions and ecosystems \citep{boakesDistortedViewsBiodiversity2010,meyerGlobalPrioritiesEffective2015,sierra2025divshift,nowosad2026navigating}. Assessing model performance or comparing structural properties across ecosystems and biomes requires dataset selection and stratification that account for the most frequently sampled regions, vegetation types, and acquisition methods \citep{cherif2025greenhyperspectra}. Gaps in geographic and ecological coverage can guide targeted data collection and contributions, including efforts to locate existing datasets that have not yet been shared.

\FloatBarrier
\subsection{Community, governance, and sustainability}
\label{sec:community_governance_sustainability}

\begin{figure}[H]
  \centering
  \resizebox{\linewidth}{!}{\hyphenpenalty=10000
\exhyphenpenalty=10000
\begin{tikzpicture}[
  x=1cm, y=1cm,
  font=\sffamily\footnotesize,
  hub/.style={draw=none, rounded corners=5pt, line width=0.9pt,
    fill=ftPlatform, align=center, text width=3.6cm, minimum height=1.6cm,
    inner xsep=6pt, inner ysep=5pt, text=ftInk},
  comm/.style={draw=none, rounded corners=5pt, line width=0.75pt,
    fill=ftField, align=center, text width=3.4cm, minimum height=1.5cm,
    inner xsep=5pt, inner ysep=4pt, text=ftInk},
  give/.style={-{Latex[length=2.2mm]}, line width=0.9pt, draw=ftLine},
  planned/.style={-{Latex[length=2.0mm]}, line width=0.65pt, draw=ftLine,
    dash pattern=on 3pt off 2pt},
  alab/.style={font=\sffamily\scriptsize, text=ftMuted, fill=white,
    inner sep=1.5pt, align=center}
]
\input{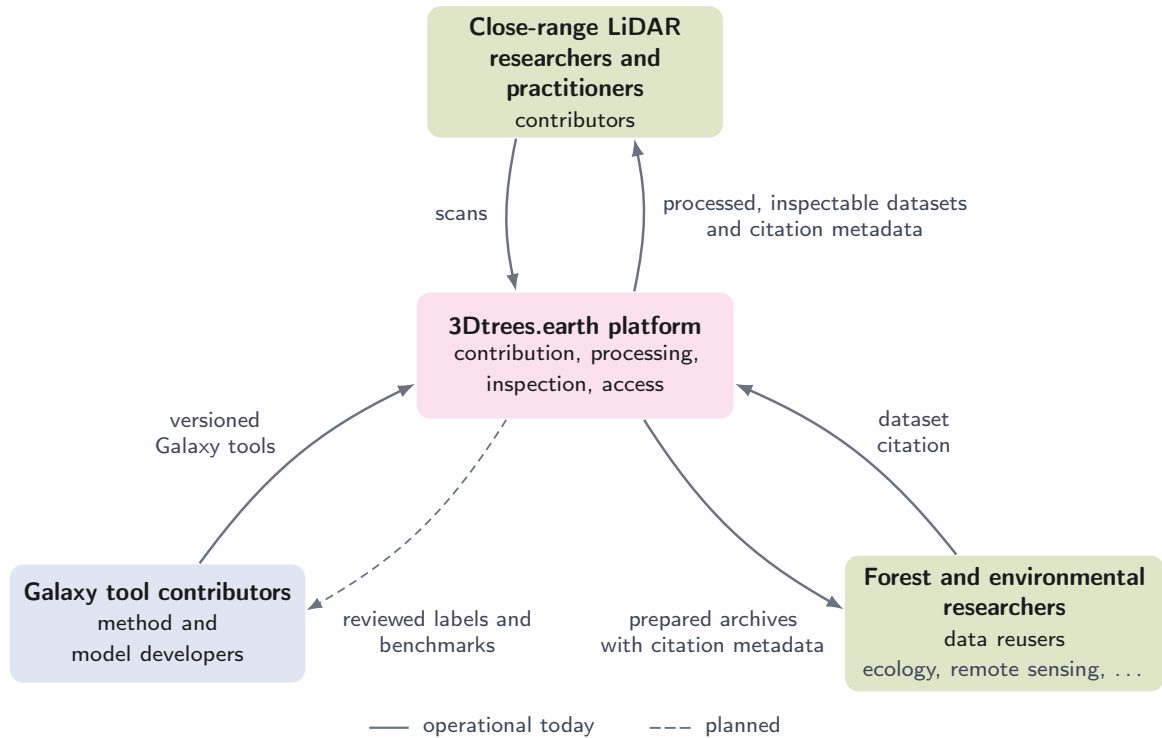}

\node[hub] (db) at (0, 0)
  {\textbf{3Dtrees.earth platform}\\
   {\scriptsize contribution, processing,\\
    inspection, access}};

\node[comm] (prod) at (0, 3.6)
  {\textbf{Close-range LiDAR researchers and practitioners}\\
   {\scriptsize contributors}};
\node[comm, fill=ftWorkflow] (meth) at (-5.3, -3.4)
  {\textbf{Galaxy tool contributors}\\
   {\scriptsize method and model developers}};
\node[comm, text width=3.7cm] (reuse) at (5.45, -3.4)
  {\textbf{Forest and environmental researchers}\\
   {\scriptsize data reusers}\\
   {\scriptsize\color{ftMuted}ecology, remote sensing, \ldots}};

\draw[give] ([xshift=-0.75cm]prod.south) to[bend right=12]
  node[alab, left=5pt]{scans} ([xshift=-0.75cm]db.north);
\draw[give] ([xshift=0.75cm]db.north) to[bend right=12]
  node[alab, right=5pt]{processed, inspectable datasets\\and citation metadata}
  ([xshift=0.75cm]prod.south);

\draw[give] (meth.55) to[bend left=14]
  node[alab, pos=0.45, above left=2pt]{versioned\\Galaxy tools} (db.190);
\draw[planned] (db.223) to[bend left=14] (meth.5);
\node[alab, fill=none] at (-1.75, -3.5) {reviewed labels and\\benchmarks};

\draw[give] (db.-43) to[bend right=14] (reuse.175);
\node[alab, fill=none] at (1.75, -3.5) {prepared archives\\with citation metadata};
\draw[give] (reuse.125) to[bend right=14]
  node[alab, pos=0.45, above right=2pt]{dataset\\citation} (db.-10);

\node[font=\sffamily\scriptsize, text=ftMuted] at (0, -4.7)
  {\textcolor{ftLine}{\rule[0.12em]{0.5cm}{0.8pt}}\; operational today
   \qquad
   \textcolor{ftLine}{\rule[0.12em]{0.14cm}{0.7pt}\,\rule[0.12em]{0.14cm}{0.7pt}\,%
   \rule[0.12em]{0.14cm}{0.7pt}}\; planned};

\end{tikzpicture}}
  \caption[Exchanges between the platform and its communities]{The 3Dtrees.earth platform as the connective element between practitioners, \textit{Galaxy} tool contributors, and researchers. Practitioners supply scans and receive processed, inspectable datasets with citation metadata. \textit{Galaxy} tool contributors, who develop methods and models, supply versioned tools; the reviewed labels and benchmarks they would receive in return are planned. Researchers receive download packages with citation metadata and return attribution through dataset citation. In practice, such communities often overlap, and the core team, which operates the platform and reviews contributions, is not shown.}
  \label{fig:community_hub}
\end{figure}

3Dtrees.earth depends on a broad, overlapping community of practitioners, method and model developers, and forest and environmental researchers (\autoref{fig:community_hub}). Practitioners contribute and manage scans, retain control over access, and receive processed, inspectable datasets with citation metadata. Method and model developers contribute tools for segmentation, species classification, or structural analysis, which can be versioned in \textit{Galaxy} and applied to datasets after technical review (Section~\ref{sec:galaxy_processing_outputs}). Researchers discover public or view-only datasets together with the context needed to select, interpret, and cite them, and return attribution through dataset citation. Deeper collaboration, including co-authorship on dataset and analysis papers, can thus be arranged between researchers with shared interests.

Sustaining these exchanges requires a core team, institutional responsibility, clear governance, and long-term infrastructure support. The core team maintains the workflows, reviews contributed datasets, and operates, develops, and maintains the platform. \autoref{tab:system_status} summarizes implemented and planned platform capabilities.

The University of Freiburg is currently responsible for operating 3Dtrees.earth. A governance board brings together research and infrastructure teams from the University of Freiburg and GFZ Helmholtz Centre for Geosciences, with broader contributor and partner representation planned. Quarterly open town-hall meetings involve the wider community as shared standards and policies develop. Further information on governance and community participation is provided on the platform's governance page\footnote{\url{https://3dtrees.earth/about/governance}}.

Sustaining open access requires protecting shared storage and processing capacity against excessive or abusive use. The platform's terms of use\footnote{\url{https://3dtrees.earth/legal/terms}} define contributor responsibilities, acceptable behavior, and procedures for reporting violations. Account-based download allowances ensure platform infrastructure stability, while safeguards include automated checks during registration to discourage bulk account creation and circumvention of these allowances. Dataset licenses govern subsequent reuse, including machine-learning applications.

Infrastructure support currently rests on two substantial in-kind contributions. The University of Freiburg Computing Center provides durable storage for original point clouds, metadata, and processed products, while the publicly funded Galaxy Europe infrastructure provides scalable computing through a partnership.

\begin{table}[t]
  \centering
  \caption[Implemented and planned platform capabilities]{Implemented and planned
  capabilities of 3Dtrees.earth as of 23 October 2026.}
  \label{tab:system_status}
  \small
  \begin{tabularx}{\linewidth}{@{}Xl@{}}
    \toprule
    \textbf{Capability} & \textbf{Status} \\
    \midrule
    Upload and structured metadata collection & Implemented \\
    Dataset access settings (public / view-only / private) & Implemented \\
    Collaborative data sharing and management & Implemented \\
    DOI metadata lookup and ORCID identifier linking & Implemented \\
    Contributor license selection and inclusion in download packages & Implemented \\
    Storage of original point clouds, processed point clouds, and predictions & Implemented \\
    \textit{Galaxy} workflow execution and processing provenance & Implemented \\
    Standardization and automated technical checks & Implemented \\
    Tree-instance segmentation and conditional species prediction & Implemented \\
    Web-based inspection and internal dataset audit & Implemented \\
    Raster products and single-tree structural metrics & Implemented \\
    Dataset download packages with citation metadata & Implemented \\
    Dataset and collection releases through FreiData with DOI assignment & Implemented \\
    \midrule
    Stand-level structural analysis and inventory summaries & Planned \\
    Research-object export (e.g.\ RO-Crate) & Planned \\
    Public presentation of dataset audit results & Planned \\
    User reporting and model-assisted label correction & Planned \\
    SpatioTemporal Asset Catalog (STAC)-compatible discovery and search & Planned \\
    Command-line access for dataset upload, download, and processing & Planned \\
    Community-curated datasets and benchmark releases & Planned \\
    \bottomrule
  \end{tabularx}
\end{table}

\FloatBarrier

\section{Outlook}
\label{sec_outlook}

3Dtrees.earth is more than a repository of datasets and a library of methods. Because the two are held together, and because every dataset remains connected to its acquisition context, processing history, and the methods applied to it, the platform can support work that neither a data archive nor a toolbox could support alone. The same observations can be revisited as methods improve, compared across sensors and forest types, and linked to observations at much larger scales. The following subsections explore these directions, including programmatic access, curated dataset releases, model benchmarking, reference data for Earth observation, and long-term records of forest structure.

\subsection{Programmatic access and curated dataset releases}
\label{sec:outlook_access_publication}

The platform already supports dataset downloads and citable releases through FreiData (Section~\ref{sec:dataset_access_publication}), but working with tens or hundreds of datasets requires ways to identify suitable data and automate their retrieval and processing. Planned developments address structured discovery, programmatic access, and scientific curation of dataset selections.

A planned SpatioTemporal Asset Catalog (STAC) will expose dataset and product metadata for structured queries, allowing users to assemble multi-dataset selections by scientific criteria rather than browsing dataset by dataset. Queries will cover original point clouds and available processing products, including standardized and prediction-enriched point clouds, raster products, and single-tree structural metrics. Extensions such as \textit{topo4d} could describe temporal relationships and alignment information for repeated acquisitions \citep{wangTopo4dTopographic4D2026}.

A command-line interface (CLI) is being developed to support scripted platform operations, beginning with dataset upload. Planned extensions include download, processing, and reprocessing while respecting the access conditions of each dataset. Existing query and exchange approaches for individual-tree point clouds provide useful models for research-oriented data access \citep{hoflePytreedbLibraryPoint2023}. Integrating spatially selective reads into these access tools would allow analysis software and machine-learning data loaders to stream selected regions from standardized or prediction-enriched point clouds without downloading the complete files.

The existing release mechanism also provides a basis for assembling citable datasets and collections around specific scientific questions. Potential features include community-built benchmark datasets for model improvement, data from scientific networks such as biodiversity experiments, and domain-specific campaigns such as disturbance events. Developing application-specific criteria for spatial resolution, geometric accuracy, and completeness would help researchers select datasets suited to their questions.

\subsection{Model evaluation and curated benchmarks}
\label{sec:outlook_methods_benchmarks}

Shared datasets and common evaluation procedures allow methods to be compared on the same basis \citep{linesAIApplicationsForest2022,CHERLET2026230,sharma2026deadtrees}. FOR-instance illustrates this: its manually annotated UAV laser scanning data come with fixed development and test subsets and guidelines for evaluating semantic and individual-tree segmentation methods \citep{pulitiFORinstanceUAVLaser2023}. FOR-species20K provides a complementary example for species classification: assembled collaboratively from close-range LiDAR point clouds and released with baseline results for seven models, it provides a common reference for evaluating tree-species classifiers \citep{pulitiBenchmarkingTreeSpecies2025}. Planned benchmark releases from 3Dtrees.earth would extend this approach to further tasks.

The platform already provides a basis for these comparisons. Its modular, versioned tools can process the same datasets under recorded conditions, with each dataset retaining its metadata and processing history (Section~\ref{sec:galaxy_processing_outputs}). \textit{SegmentAnyTree} \citep{wielgoszSegmentAnyTreeSensorPlatform2024}, \textit{ForestMamba} \citep{nguyenForestMambaSparse2026}, and \textit{RayCloudTools} \citep{devereuxRayExtractFastScalable2026} already run on the same datasets, and their tree-instance predictions can be downloaded, visualized, and compared side by side.

Reference labels are needed to turn these comparisons into evaluations of model performance. 3Dtrees.earth already contains annotated datasets, including FOR-instance and TreeScanPL10k \citep{pulitiFORinstanceUAVLaser2023,sterenczak_2026_19127709}. Harmonizing their label definitions and formats would provide reference data for training and evaluation. Additional annotation efforts could extend this resource to datasets that currently lack labels. However, annotating platform-hosted data in desktop tools such as \textit{CloudCompare} \citep{CloudCompare} requires users to download point clouds, edit them locally, and upload and reintegrate the resulting labels.

We therefore plan to integrate labeling tools into the platform's web-based viewer, allowing users to annotate unlabeled point clouds, refine existing reference labels, and correct model predictions directly in the browser. Spatial clipping and selection tools would allow reviewers to merge fragments of the same tree and separate incorrectly joined trees. Promptable correction with tools such as \textit{SelectAnyTree} \citep{nguyenSelectAnyTreePromptable2026} would allow reviewers to refine a predicted tree using positive and negative point prompts. A complementary semi-automatic approach could cluster learned point embeddings into candidate tree instances for reviewers to merge or split. Planned user reporting would direct implausible predictions into this review process. Contributed labels would undergo review before inclusion in a benchmark.

As reviewed annotations accumulate, they would support fixed, versioned, and citable benchmark releases for defined tasks. Each release would specify the selected datasets, reviewed labels, training, validation, and test splits, evaluation metrics, baseline model predictions or workflows, and the relevant workflow and tool versions.

\subsection{Reference data for Earth observation model development and validation}
\label{sec:outlook_reference_data}

EO missions map forest attributes across regions and continents, with spatial and temporal coverage beyond that of close-range scanning. Whether derived from optical, radar, or spaceborne LiDAR observations, their products represent forest structure at coarser scales and require calibration and validation against finer-scale ground measurements, which remain scarce \citep{kattenborn2019uav,kattenbornReviewConvolutionalNeural2021,schiefer2023uav}. Crowdsourced close-range scans resolve three-dimensional forest structure at tree and plot scales and can therefore provide reference data for the development and validation of EO models and products \citep{liangCloseRangeRemoteSensing2022,labriereForestBiomassReference2023}.

Spatially referenced scans can be matched to airborne or spaceborne footprints when location, acquisition date, spatial extent, and scale align closely enough for the intended comparison. Datasets from 3Dtrees.earth could then support the development and evaluation of EO models and products for canopy height, structural complexity, biomass, species composition, mortality, or disturbance \citep{liangCloseRangeRemoteSensing2022,networkGlobalUnderstandingTree2025,schieferLargescaleRemoteSensing2025,Mosig2026_SubPixelDisturbance}. Translating LiDAR observations into EO reference layers depends on the target property: geometric variables, such as canopy height or canopy gap fraction, can be derived directly from the scans. Species, mortality, and disturbance require appropriate labels or independent observations of the attribute or event being assessed. Biomass and carbon estimates require additional information, such as wood density and, depending on the method, allometric relationships \citep{caldersNondestructiveEstimatesAboveground2015,labriereForestBiomassReference2023}.

Because the original scans are retained, reference layers can be recomputed as methods improve, allowing the same observations to serve EO questions posed after acquisition, including those concerning properties not yet routinely derived from close-range LiDAR \citep{liangCloseRangeRemoteSensing2022}. Planned programmatic access and curated multi-dataset releases would make purpose-built selections available at collection scale. Reference layers could also be tailored to a mission's requirements by reprojecting and resampling them to its coordinate reference system and spatial resolution.

Because 3Dtrees.earth is crowdsourced, its coverage is uneven. Some regions are densely represented, whereas others remain sparsely covered (Section~\ref{sec:platform_scope}). The catalog and its map can help identify where new acquisitions would improve coverage for calibrating and validating EO products \citep{networkGlobalUnderstandingTree2025,schieferLargescaleRemoteSensing2025}. Its usefulness can grow through both methodological advances that extract more from existing scans and targeted contributions that fill spatial gaps.

Established in situ resources, including national forest inventories, could fill some of these gaps, but access restrictions and predefined sets of measured variables can limit their use as scientific questions change \citep{labriereForestBiomassReference2023}. Complementing these resources with reusable close-range observations and community contributions could expand the reference data available for EO applications and the broader study of forest structure \citep{ehbrecht2021global,atkins2023integrating}.

\subsection{Translating multitemporal scans to digital twins and models}
\label{sec:outlook_digital_twins}

Point clouds and derived products in 3Dtrees.earth already have a temporal component through their acquisition dates. However, observations of the same site are not yet necessarily geometrically or thematically consistent through time. Scans may not be spatially aligned, while derived variables, tree instances, or labels may differ because of changes in sensors, acquisition designs, or processing methods. Forests themselves also change on short to seasonal and longer timescales through wind-induced branch movement, phenology, growth, and mortality. These changes can complicate co-registration and must be distinguished from differences introduced by acquisition or processing \citep{weiserAdvancingVegetationMonitoring2026}. Geometrically and thematically consistent temporal records are a prerequisite for forest monitoring, growth and yield modeling, and digital twins.

For temporal collections, future 3Dtrees.earth workflows will co-register repeated scans and quantify geometric change using established 4D point cloud methods \citep{andersPy4dgeoOpenSourceScientific2026}, while documenting thematic differences among acquisitions. Derived products will include explicit temporal information, persistent site and tree identities, and links to their source scans and processing runs. The result is a four-dimensional record combining three-dimensional structure with time. At the individual tree level, such records could track growth, crown development, and changes in other structural attributes, while at the stand level they could reveal changes in tree density, recruitment, mortality, and disturbance.

Coherent multitemporal records can support forest modeling and digital twins. Once tree-level inventory products, such as validated positions, species, height, stem diameter, and crown dimensions, can be derived from 3Dtrees.earth scans, they could provide initial conditions for forest growth and dynamics models, supporting projections of growing-stock increment. A concrete integration pathway is being developed through the Forest Digital Twin Database \citep{sperlich_digital_2026}, which assigns persistent identities to individual trees and links measured and simulated forest states through time within site-specific management scenarios. These records can be analyzed and visualized in web dashboards or 3D applications. Prototype interfaces already prepare forest inventories for the \textit{SILVA} growth model \citep{pretzsch2002single} and write projected tree dimensions back to the database.

Linked to growth models in this way, multitemporal LiDAR-derived inventories could provide spatially and temporally explicit initial conditions for scenario generation, such as forest growth under climate projections following different socio-economic pathways \citep{egerer2025forestation, gomez2025climate}. The same records, combined with environmental data and site conditions, could also support vegetation model calibration and validation. The spatial and structural detail that close-range LiDAR provides is particularly suited to questions about how tree structure shapes microclimate, competition, and biodiversity \citep{blanchard2023uav,wang2025influence,schnabelTreeDiversityTemperature2025,kroner2026disentangling}.

Temporally and spatially coherent LiDAR inventories could also be linked to past or live measurement streams, including dendrometer records, ecosystem fluxes, and environmental variables. Aligning these observations in space and time would connect changes in tree and stand structure with physiological responses and environmental conditions. The Forest Digital Twin Database provides a potential route for storing these data alongside measured and projected forest states and making them accessible through a common interface \citep{sperlich_digital_2026}.

\begin{figure*}[tbp]
  \centering
  \includegraphics[width=\textwidth]{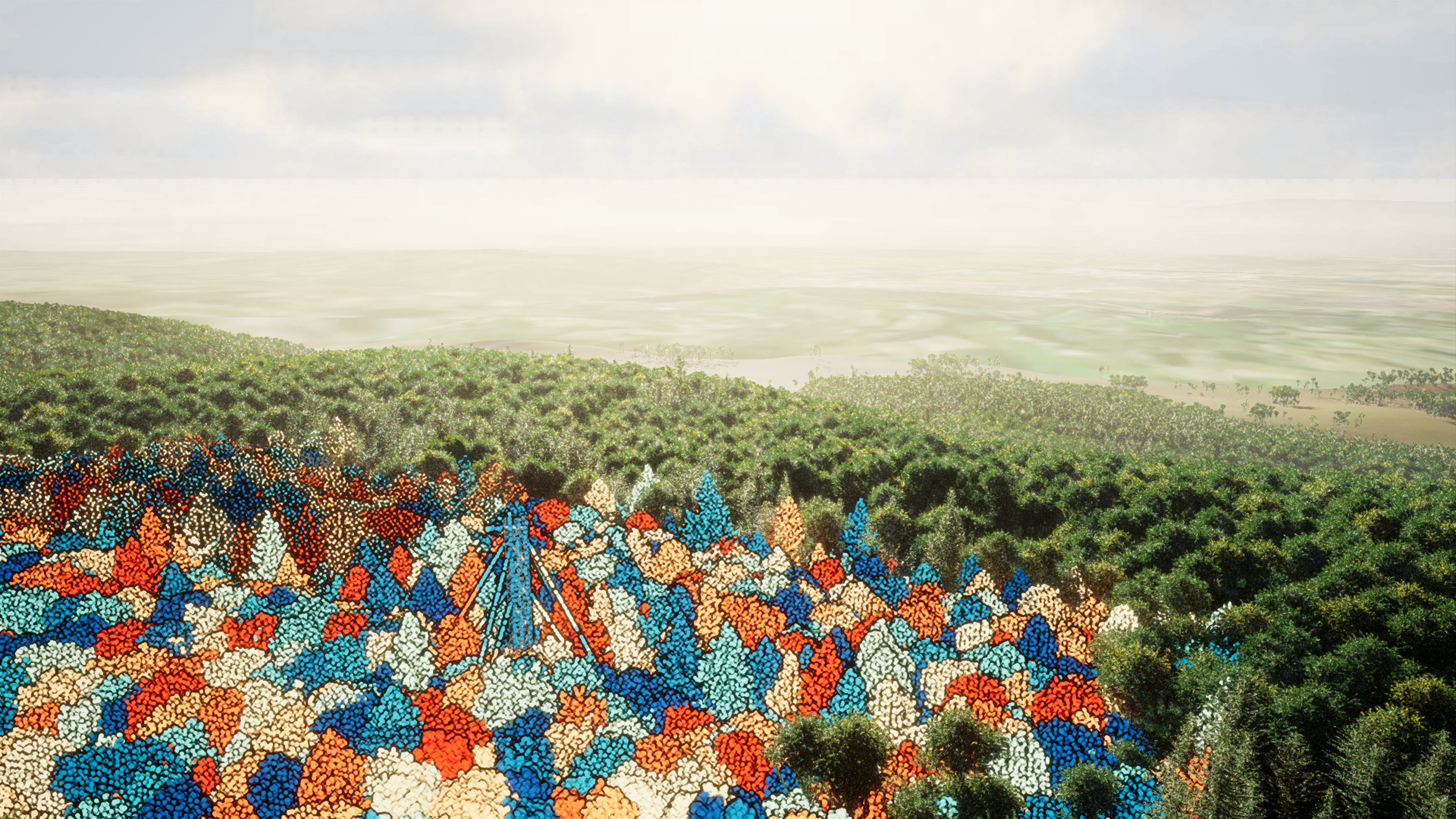}
\caption[Point clouds and virtual forest visualization]{Illustration combining tree-instance-colored point clouds in the foreground with a virtual forest scene rendered in \textit{Unreal Engine}. The contrasting colors distinguish individual tree instances. Such environments can support spatially explicit exploration of forest structure and scenario-based projections. Direct transfer of 3Dtrees.earth processing products into this environment is planned.}
  \label{fig:unreal_forest_environment}
\end{figure*}

LiDAR-derived inventories can also be translated into geometrically explicit virtual forests, providing the basis for photorealistic and immersive representations of measured or projected forest states for research, communication, and education. In an existing workflow, \textit{GrowPy} uses measured or projected tree dimensions and positions to parameterize individual trees and passes them to \textit{The Grove}, a procedural tree-generation engine that creates spatially positioned three-dimensional tree meshes \citep{sperlich_growpy_2026}. These meshes can be explored in \textit{Unreal Engine} (\autoref{fig:unreal_forest_environment}) or web viewers, and can provide virtual forest scenes for \textit{HELIOS++} simulations, which generate synthetic, annotated LiDAR point clouds \citep{winiwarter2022virtual,schaferGeneratingSyntheticLaser2023}. With suitable canopy optical properties and illumination conditions, these scenes could also support evaluation of radiative-transfer models and remote-sensing retrieval methods. The Radiation Transfer Model Intercomparison (RAMI) initiative illustrates this approach by comparing model simulations for shared three-dimensional canopy scenarios \citep{lanconelliRAMIV2025}. Throughout, the generated scenes and simulated point clouds retain links to the measurements or model projections they were built from. \autoref{fig:digital_twin_data_flow} in \autoref{app:digital_twin} summarizes this flow, from 3Dtrees.earth and associated data streams through the Forest Digital Twin Database to visualization and synthetic data.

\subsection{A gateway to macroecological research and digital forest archives}

As community participation in 3Dtrees.earth grows, contributions can increase the volume, diversity, and geographic coverage of harmonized, high-resolution 3D forest data. Curation and validation can improve the quality and suitability of this resource for scientific reuse. This cumulative resource can support analyses that move beyond individual sites and case studies, providing the distributed data foundation for investigating forest structure and dynamics across regional to global scales. Community engagement therefore does more than accelerate methodological development: it expands the scope of ecological questions that can be addressed.

Forests are complex three-dimensional systems whose architecture reflects evolutionary history, species interactions, competition, land use, and climate \citep{pan2013structure,garnier2016plant}. Yet the scarcity of comparable structural measurements across regions has constrained ecological, biogeographic, and macroecological analyses of forest variability \citep{ehbrecht2021global,atkins2023integrating}.

A geographically distributed and standardized collection of forest LiDAR observations can support systematic comparisons of forest structure, state, and functioning across ecosystems, biomes, and land-use gradients. By providing standardized point clouds together with acquisition metadata, 3Dtrees.earth helps researchers select suitable datasets and account for differences in acquisition methods, point density, and occlusion. Shared storage, structured metadata, and common processing workflows could support coordinated analyses across sites within research networks such as the Tree Diversity Network (TreeDivNet), the Integrated Carbon Observation System (ICOS) \citep{rebmannICOSFluxStation2018}, and the Integrated European Long-Term Ecosystem, Critical Zone and Socio-Ecological Research infrastructure (eLTER) \citep{mollenhauerLongTermMonitoring2018}. Where observations coincide in space and time, LiDAR-derived structural records could be combined with these networks' field measurements to investigate relationships between forest structure, biodiversity, carbon exchange, and long-term ecosystem change.

Beyond biogeographic and macroecological research, the same distributed and increasingly multitemporal collection also serves as a digital archive of forest structure. This archival role is becoming more urgent. Global forest area declined by a net 178 million hectares between 1990 and 2020 \citep{faoGlobalForestResourcesAssessment2020}, while land-cover statistics capture neither structural degradation nor the loss of ecological complexity \citep{granthamAnthropogenicModificationForests2020, hansenTowardMonitoringForest2021}. Some ecologically distinctive and old-growth forests are being altered in ways that may not be reversible for centuries \citep{gibsonPrimaryForestsIrreplaceable2011,lindenmayerGlobalDeclineLarge2012,watsonExceptionalValueIntact2018}. Close-range LiDAR documents forest architecture, vertical complexity, and fine-scale heterogeneity, providing detailed structural records for long-term ecological research and digital heritage. Where access conditions allow, these records can also support virtual forest visits and immersive educational applications. 3Dtrees.earth can thus support standardized, high-resolution documentation of forest structure across regions and forest types.

\section{Conclusion}
3Dtrees.earth connects close-range LiDAR observations of trees and forests with acquisition metadata, processing methods, visualization, and data access. These connections reduce the work required to move from a contributed survey to scientific reuse. Retaining the original observations and their processing context allows datasets to be revisited as methods improve and new scientific questions arise.

Continued contributions can extend the geographic and temporal coverage of close-range LiDAR observations and support research beyond individual surveys. Realizing this potential requires sustained maintenance, community participation, and curation alongside shared infrastructure such as Galaxy Europe. By keeping observations, methods, and context connected, 3Dtrees.earth provides a basis for collaborative method development, cross-site synthesis, and long-term study of tree and forest structure.

\section*{Author contributions}
J.V.-J. and K.G. contributed equally to this work. J.V.-J. led platform development, with contributions from D.Lu. K.G. led analytical workflow development and figure preparation. M.Sp. contributed Figures~\ref{fig:unreal_forest_environment} and~\ref{fig:digital_twin_data_flow}. T.Ka. and J.Fr. contributed to scientific design, methodology and supervision. K.G. wrote the initial manuscript draft, and J.V.-J. and K.G. led its revision with substantial input from D.Lu., T.Ka. and J.Fr. The wider author team contributed datasets and collections, analytical methods, research infrastructure, data curation, and manuscript review and revision.

\section*{Acknowledgements}

T.Ka., D.Lu., J.V.-J., K.G. and J.Fr. acknowledge support from the Deutsche Forschungsgemeinschaft (DFG, German Research Foundation) under Germany's Excellence Strategy through EXC 3127 Future Forests (project no. 533786343), the Emmy Noether project PANOPS (project no. 504978936), NFDI4Earth (project no. 460036893), 3Dtrees.dat (project no. 575836471), and OpenS2Hub (project no. 593998785), as well as from the Eva Mayr-Stihl Foundation.

T.Ka., K.G., J.Sei. and A.Re. acknowledge support from the Deutsche Forschungsgemeinschaft (DFG, German Research Foundation) through the project LeafH2O, a UAV-mounted dual-wavelength LiDAR for leaf water content retrieval (project no. 541018379).

R.V. acknowledges support from the Circular Bio-based Europe Joint Undertaking and its members under grant agreement No. 101157488 (SingleTree). Funded by the European Union. Views and opinions expressed are however those of the author(s) only and do not necessarily reflect those of the European Union or CBE JU. Neither the European Union nor the CBE JU can be held responsible for them.

A.Gö. acknowledges funding from the German Research Foundation (DFG) under Project-ID 459819582 (SFB 1537). N.E. acknowledges support from the Deutsche Forschungsgemeinschaft (DFG, German Research Foundation) through the Gottfried Wilhelm Leibniz Prize (grant no. Ei 862/29-1) and the German Centre for Integrative Biodiversity Research (iDiv) Halle-Jena-Leipzig, funded by the DFG (grant no. FZT 118, 202548816).

K.C., W.C., W.A.J.V.d.B. and Z.C. were funded by the European Union (ERC-2021-STG, grant agreement No. 101039795). Views and opinions expressed are however those of the author(s) only and do not necessarily reflect those of the European Union or the European Research Council Executive Agency. Neither the European Union nor the granting authority can be held responsible for them.

B.B., L.L. and J.Wi. acknowledge funding from the FORWARDS project (grant agreement no. 101084481). B.B., L.L., J.Wi., R.V., M.Hero., B.Gi. and G.D.S. acknowledge funding from the NextGenCarbon project (grant agreement no. 101184989).

Z.S. and C.Mor. acknowledge the HUMAX project, which is supported by funds of the Federal Ministry of Agriculture, Food and Regional Identity (BMLEH) based on a decision of the Parliament of the Federal Republic of Germany via the Fachagentur Nachwachsende Rohstoffe e.V. (FNR) under the Climate protection measure Humus, grant number 2224HA005.

F.P. and E.K. acknowledge funding from Fondazione Cassa di Risparmio di Padova e Rovigo (CARIPARO), which supported close-range sensing data acquisition in San Vito di Cadore (Dolomites). E.R.L., H.O. and S.W.D.G. were funded by a UKRI Future Leaders Fellowship awarded to E.R.L. (MR/T019832/1 and MR/Y033981/1).

Datasets produced by M.Do., A.G.-S., T.Kl. and M.Sa. originated from projects funded by the Deutsche Forschungsgemeinschaft (DFG, German Research Foundation; project number KL 2265/10-1), the Hessischer Biodiversitätsforschungsfonds of the Hessian Agency for Nature Conservation, Environment and Geology (project numbers Z1--15 c 02.01 and Z1--15 c 02.03), and the Hessian Ministry of Higher Education, Research and the Arts of the Federal State of Hesse, Germany (grant ID 408/05.017 - (0003) f)). M.Do. and R.Be. acknowledge funding from INTERREG Deutschland-Nederland (project number 143081) for the development of a ULS system.

J.A.G. acknowledges funding from the European Union (ERC, ADAPTA, 101231095), the Natural Environment Research Council (NE/T011084/1 and NE/Z504191/1), the Royal Society (RG\textbackslash R1\textbackslash 251370), and the Leverhulme Trust (RPG-2024-342).

S.Kr. acknowledges funding for instruments from the Ministerium für Wissenschaft, Forschung und Kultur (grant/award no. F221-08-AWI/001/002) and funding for point cloud processing to the publication stage from the German Research Foundation (DFG; grant/award no. 563037533, Assessing stability of topographic impacts on northern boreal forest carbon allocation across the Tundra-Taiga Ecotone amid climate crisis, BToBE).

We sincerely thank Nishiawakura Village, Okayama, Japan, Hyakumori Inc., Nishiawakura Village, Okayama, Japan, and Fuji Iyashinomori Woodland Study Center, The University of Tokyo, Yamanashi, Japan, for kindly providing access to forest sites used for data collection in this study.

A.Gr., J.R.H., M.P.E. and G.Sh. acknowledge funding from the Natural Environment Research Council through the Envision Doctoral Training Partnership (NE/L002604/1). A.Gr. and G.Sh. also acknowledge funding from the Wales Biodiversity Partnership through the Biodiversity and Ecosystem Evidence and Research Needs (BEERN) Programme.

A.M.P. acknowledges support from the Slovenian Research and Innovation Agency (ARIS) through the research programme P4-0107 (Forest Biology, Ecology and Technology) and the infrastructure programme I0-0012 (Infrastructural Programme of the Slovenian Forestry Institute).

UAS data collection for Pietrari, Romania, was supported by the LEAD-Landslide project (PN-IV-PCB-RO-MD-2024-0375; contract 40PCBROMD/01.09.2025), funded through the Romanian National Plan for Research, Development and Innovation (PNCDI IV) and administered by the Executive Agency for Higher Education, Research, Development and Innovation Funding (UEFISCDI). UAS data collection for Iron Gate Natural Park received funding from the European Union's Horizon Europe research and innovation programme under grant agreement no. 101181408 (BioClima: Improving Monitoring for Better Integrated Climate and Biodiversity Approaches, Using Environmental and Earth Observations).

Datasets collected at Wakehurst, Kew, UK, are part of the Nature Unlocked programme at Royal Botanic Gardens, Kew, funded under the Nature Returns programme through the UK government's Shared Outcomes Fund, funded by HM Treasury and co-sponsored by DEFRA and DESNZ. Nature Returns is delivered in partnership with Natural England, the Environment Agency, the Forestry Commission, the University of Sussex and local partners.

Datasets produced by C.Rü. and A.El. were acquired with a CHCNav RS10 handheld laser scanner purchased within the project ``Cross-border forest monitoring in the Bohemian Switzerland/Saxon Switzerland National Park'' (TransForMon, project no. 100736674), funded by the Interreg Saxony--Czech Republic 2021--2027 programme and co-funded by the European Union through the European Regional Development Fund (ERDF).

E.P., M.Ca. and Y.W. acknowledge support for the collection and processing of the LiPhe dataset at the Finnish Geospatial Research Institute of the National Land Survey of Finland (NLS-FGI) from the Research Council of Finland through ForestCycles (grant no. 334060), LS-HYDRO (grant no. 356137), and HPC-Carbon (grant nos. 359203 and 359204), and from the European Union's Horizon programme through TRANSFORMIT (grant no. 101135263).

P.M.E. and W.O. acknowledge funding through the LabForest project from the Bundesministerium für Forschung, Technologie und Raumfahrt (BMFTR; grant no. 033L307A).

T.A.R. acknowledges funding from ANID FONDECYT Iniciación Grant No. 11250090.

D.A. acknowledges funding from the AUDI Environmental Foundation through the project ``Spatial-temporal growth reaction of oaks to growing and site conditions'' (No. 5104146).

K.Krá. and M.Kr. acknowledge funding from the Institutional Support of LRI (IP-00027073).

Datasets provided by R.J., C.Sh. and C.I.B.W. originate from the project KI-Recover, funded by the Federal Ministry for the Environment, Climate Action, Nature Conservation and Nuclear Safety (BMUKN, grant number 67KIA4059A).

\section*{Competing interests}
The authors declare no competing interests.

\section*{Data availability}
Datasets are accessible through \url{https://3dtrees.earth} under their individual access conditions and licenses. Public datasets can be downloaded by registered users; downloads of view-only and private datasets require permission. Dataset records provide contributor attribution and links to existing source publications or repository records where available.

\section*{Code availability}
The application repository is not publicly available because it contains non-public information shared among contributors and internal operational material requiring confidentiality. Publicly available processing tools and workflows are cited separately.

\section*{Declaration of generative AI and AI-assisted technologies in the manuscript preparation process}
During the preparation of this work, the authors used OpenAI tools (ChatGPT and Codex) and Claude (Anthropic) to assist with drafting and revising the manuscript, improving its language and structure, checking consistency, and writing and debugging code for the platform, analyses, and figure preparation. Generative AI was also used to create selected illustrative elements in Figure~\ref{fig:3dtrees_main}; these are schematic illustrations, not measured data or analytical results. The authors reviewed and edited the outputs as needed and take full responsibility for the content of the published article.

\begingroup
\setlength{\emergencystretch}{2em}
\printbibliography[title={References}]
\endgroup

\clearpage
\appendix
\counterwithin*{figure}{section}
\counterwithin*{table}{section}
\renewcommand{\thefigure}{\thesection.\arabic{figure}}
\renewcommand{\thetable}{\thesection.\arabic{table}}
\section{Dataset metadata}
\label{app:dataset_metadata}

\autoref{tab:dataset_metadata} summarizes the metadata collected through the dataset upload form. Required fields refer to form submission, not to a guarantee of metadata completeness or scientific accuracy. The field set can evolve with community requirements. Preprocessing categories describe steps reported by contributors; they do not constitute a reproducible record of algorithms and parameter settings. Acquisition date is recorded, but leaf-on or leaf-off state and weather conditions have no dedicated fields. Contributors can provide additional context in the description or supplementary material.

\begingroup
\small
\singlespacing
\setlength{\tabcolsep}{4pt}
\renewcommand{\arraystretch}{1.0}
\begin{longtable}{@{}>{\RaggedRight\arraybackslash}p{0.23\linewidth}>{\RaggedRight\arraybackslash}p{0.19\linewidth}>{\RaggedRight\arraybackslash}p{\dimexpr0.58\linewidth-4\tabcolsep\relax}@{}}
\caption{Dataset upload metadata in the implementation inspected in October 2026. Optional entries may be absent. Automatically extracted information is limited by the content and readability of the uploaded files.}\label{tab:dataset_metadata}\\
\toprule
Field & Entry requirement & Content \\
\midrule
\endfirsthead
\multicolumn{3}{l}{\tablename\ \thetable\ (continued)}\\
\toprule
Field & Entry requirement & Content \\
\midrule
\endhead
\bottomrule
\endfoot
Title & Conditional & Required for multi-file datasets; otherwise the filename is used when no title is supplied. \\
Description & Optional & Free-text description of the dataset or project. \\
Acquisition date & Required & Date of the survey. \\
Acquisition platform & Required & Terrestrial (TLS), mobile (MLS), or uncrewed aerial laser scanning (ULS). \\
Sensor & Optional & Instrument name, selected or entered as text. \\
Scanning pattern & Optional & Free text, for example a grid spacing or transect arrangement. \\
Prior processing & Optional & Noise filtering, ground classification, height normalization, downsampling, co-registration, outlier removal, and compression; multiple categories may be selected. \\
Coordinate reference system & Automatic, where available & Read from the point cloud; absence of a usable reference system requires an approximate location. \\
Coordinate accuracy & Required & Contributor-selected category: below 1\,m, below 10\,m, above 10\,m, or local coordinates. \\
Approximate location & Conditional & Latitude and longitude supplied when a usable coordinate reference system is missing. \\
Label or prediction mappings & Conditional & Value-to-label mappings are required when the contributor indicates that the uploaded files contain labels or predictions. This does not establish their accuracy. \\
Contributors & Required & At least one contributor's first and last name; organization and ORCID identifier are optional. \\
Funding & Optional & Funder name, with optional funder identifier, award number, and award title. \\
Linked DOI & Optional & Identifier of an existing publication or repository record; available metadata can be retrieved from that record. \\
Supplementary material & Optional & ZIP archive and an optional description, for example supporting documentation or accompanying field data. \\
Visibility & Required & Public, view-only, or private access setting. \\
Location disclosure & Optional & Setting to hide the exact location from users without the corresponding access. \\
License and rights confirmation & Required & Selected reuse license and confirmation that the contributor holds the rights to license the dataset. \\
\end{longtable}
\endgroup

\clearpage
\section{Platform architecture}
\label{app:platform_architecture}

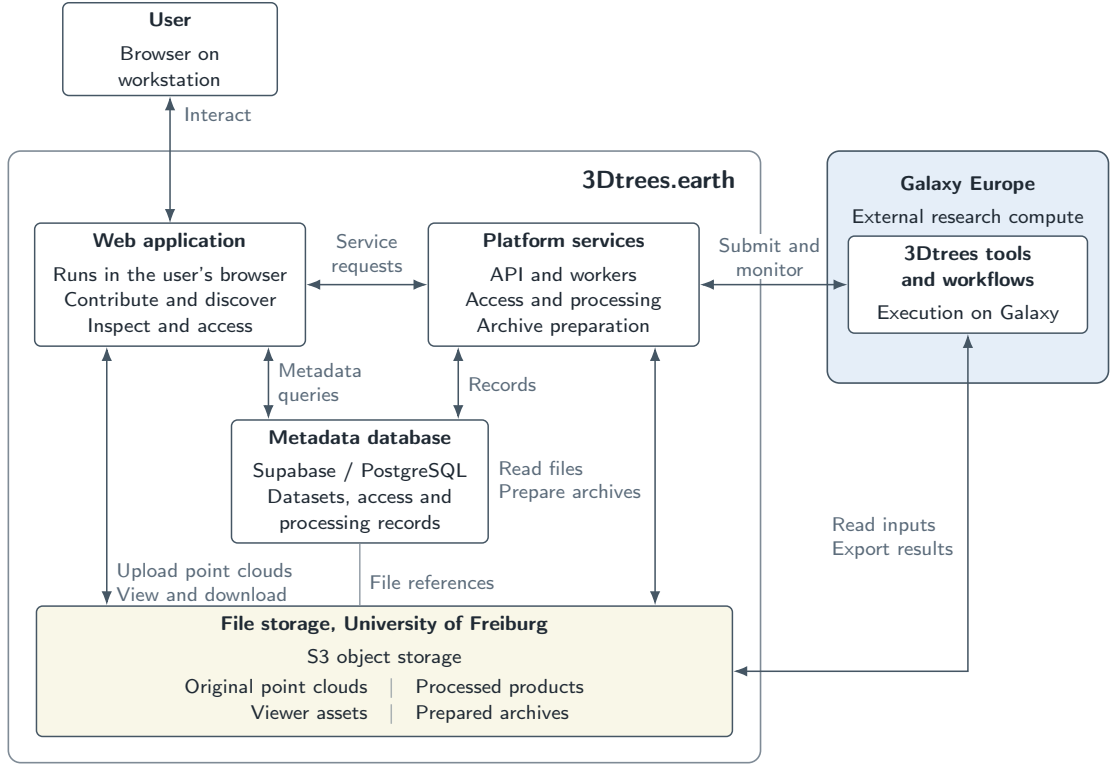
\begin{figure}[htbp]
  \centering
  \resizebox{\linewidth}{!}{\begingroup
\definecolor{dtInk}{HTML}{1F2933}
\definecolor{dtMuted}{HTML}{5B6B7A}
\definecolor{dtLine}{HTML}{7F8B96}
\definecolor{dtLineStrong}{HTML}{455664}
\definecolor{dtProcessing}{HTML}{E5EEF8}
\definecolor{dtOutput}{HTML}{F9F7E8}
\hyphenpenalty=10000
\exhyphenpenalty=10000
\begin{tikzpicture}[
  x=1cm,y=1cm,
  font=\sffamily\fontsize{8.5}{10.5}\selectfont,
  text=dtInk,
  component/.style={draw=dtLineStrong,rounded corners=2.5pt,line width=.6pt,
    fill=white,minimum width=3.8cm,minimum height=1.75cm,text width=3.5cm,
    inner xsep=5pt,inner ysep=4pt,align=center,text=dtInk},
  label/.style={font=\sffamily\fontsize{8}{9.5}\selectfont,align=left,
    text=dtMuted,fill=white,inner sep=1.5pt},
  transfer/.style={{Latex[length=2.0mm]}-{Latex[length=2.0mm]},
    line width=.62pt,draw=dtLineStrong},
  reference/.style={line width=.5pt,draw=dtLine}
]
\draw[draw=dtLine,rounded corners=5pt,line width=.65pt]
  (-.4,-.6) rectangle (10.3,8.1);
\node[anchor=north east,font=\sffamily\bfseries\fontsize{10}{12}\selectfont]
  at (10.08,7.96) {3Dtrees.earth};
\node[component,minimum width=3cm,text width=2.7cm,minimum height=1.1cm]
  (user) at (1.9,9.55)
  {\textbf{User}\\[3pt]Browser on workstation};
\draw[transfer] (user.south) -- (1.9,7.075);
\node[label,anchor=west] at (2.04,8.62) {Interact};
\node[component] (browser) at (1.9,6.2)
  {\textbf{Web application}\\[3pt]
   Runs in the user's browser\\
   Contribute and discover\\
   Inspect and access};
\node[component] (app) at (7.5,6.2)
  {\textbf{Platform services}\\[3pt]
   API and workers\\
   Access and processing\\
   Archive preparation};
\draw[draw=dtLineStrong,fill=dtProcessing,rounded corners=5pt,line width=.65pt]
  (11.25,4.8) rectangle (15.25,8.1);
\node[align=center,anchor=north] at (13.25,7.88)
  {\textbf{Galaxy Europe}\\[3pt]External research compute};
\node[component,minimum width=3.4cm,text width=3.05cm,minimum height=1.3cm]
  (galaxy) at (13.25,6.2)
  {\textbf{3Dtrees tools}\\\textbf{and workflows}\\[3pt]Execution on Galaxy};
\node[component,minimum width=3.6cm,
  text width=3.3cm,minimum height=1.7cm] (db) at (4.6,3.4)
  {\textbf{Metadata database}\\[3pt]
   Supabase / PostgreSQL\\
   Datasets, access and\\processing records};
\node[component,fill=dtOutput,minimum width=9.9cm,
  text width=9.3cm,minimum height=1.4cm] (store) at (4.95,.7)
  {\textbf{File storage, University of Freiburg}\\[3pt]
   S3 object storage\\[2pt]
   \begin{tabular}{@{}r@{\quad\textcolor{dtLine}{|}\quad}l@{}}
     Original point clouds & Processed products\\
     Viewer assets & Prepared archives
   \end{tabular}};

\draw[transfer] (browser.east) -- (app.west);
\node[label,align=center,anchor=south] at (4.7,6.27) {Service\\requests};
\draw[transfer] (app.east) -- (galaxy.west);
\node[label,align=center,anchor=south] at (10.43,6.27) {Submit and\\monitor};

\draw[transfer] ([xshift=1.4cm]browser.south)
  -- ([xshift=1.4cm]browser.south |- db.north);
\node[label,anchor=west] at (3.38,4.79) {Metadata\\queries};
\draw[transfer] ([xshift=-1.5cm]app.south)
  -- ([xshift=-1.5cm]app.south |- db.north);
\node[label,anchor=west] at (6.08,4.79) {Records};

\draw[transfer] ([xshift=-.9cm]browser.south)
  -- ([xshift=-.9cm]browser.south |- store.north);
\node[label,anchor=west] at (1.08,1.97) {Upload point clouds\\View and download};
\draw[reference] (db.south) -- (db.south |- store.north);
\node[label,anchor=west] at (4.68,1.97) {File references};
\draw[transfer] ([xshift=1.3cm]app.south)
  -- ([xshift=1.3cm]app.south |- store.north);
\node[label,anchor=east] at (8.66,3.4) {Read files\\Prepare archives};
\draw[transfer] (galaxy.south) -- (13.25,.7) -- (store.east);
\node[label,anchor=east] at (13.1,2.6) {Read inputs\\Export results};
\end{tikzpicture}
\endgroup}
  \caption[High-level platform architecture]{High-level software architecture of 3Dtrees.earth. The platform grouping identifies its components, not a common hosting location. Users interact with the web application in their browsers. Platform services coordinate access and processing; 3Dtrees tools and workflows execute on the external Galaxy Europe infrastructure. Metadata and processing records are held in Supabase/PostgreSQL, while files reside in University of Freiburg object storage. Arrows indicate interaction or request and data exchange, including authorized transfers between the web application and storage and between Galaxy and storage. The unarrowed line denotes references from metadata records to stored files, not file transfer through the database.}
  \label{fig:platform_architecture}
\end{figure}

\clearpage
\section{Digital forest twin integration}
\label{app:digital_twin}

\begin{figure*}[h]
  \centering
  \resizebox{\textwidth}{!}{\centering
\input{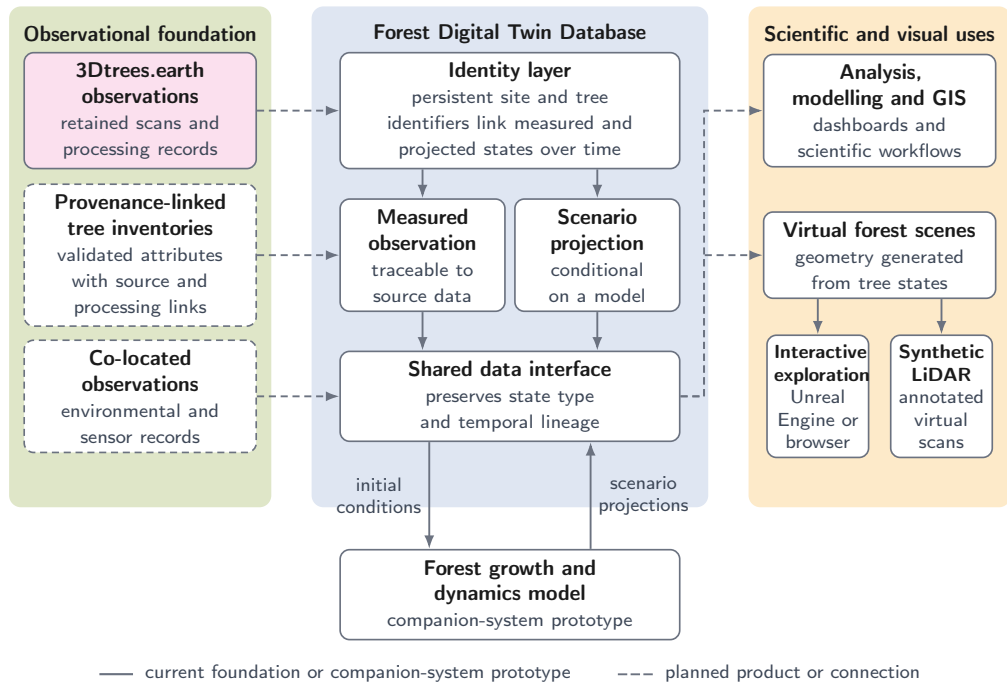}
\hyphenpenalty=10000
\begin{tikzpicture}[x=1cm, y=1cm,
  box/.style={rounded corners=4pt, fill=white, align=center, text width=3.1cm,
    minimum height=1.05cm, inner xsep=5pt, inner ysep=4pt, text=ftInk,
    font=\sffamily\footnotesize, draw=ftLine, line width=0.7pt},
  plan/.style={box, dash pattern=on 3pt off 2pt},
  half/.style={text width=1.3cm, inner xsep=3pt, font=\sffamily\scriptsize},
  sub/.style={font=\sffamily\scriptsize, text=ftMuted},
  arr/.style={-{Latex[length=2mm]}, line width=0.8pt, draw=ftLine},
  parr/.style={arr, dash pattern=on 3pt off 2pt},
  head/.style={font=\sffamily\footnotesize\bfseries, text=ftInk, align=center},
  lab/.style={font=\sffamily\scriptsize, text=ftMuted, align=center, inner sep=1pt},
]
\fill[ftField, rounded corners=5pt]    (-1.95,0.55) rectangle (1.95,8.0);
\fill[ftWorkflow, rounded corners=5pt] (2.55,0.55) rectangle (8.45,8.0);
\fill[ftStandard, rounded corners=5pt] (9.05,0.55) rectangle (12.95,8.0);
\node[head] at (0,7.6)    {Observational foundation};
\node[head] at (5.5,7.6)  {Forest Digital Twin Database};
\node[head] at (11.0,7.6) {Scientific and visual uses};

\node[box, fill=ftPlatform] (obs) at (0,6.45)
  {\textbf{3Dtrees.earth observations}\\{\scriptsize\color{ftMuted}retained scans and processing records}};
\node[plan] (inv) at (0,4.3)
  {\textbf{Provenance-linked tree inventories}\\{\scriptsize\color{ftMuted}validated attributes with source and processing links}};
\node[plan] (col) at (0,2.2)
  {\textbf{Co-located observations}\\{\scriptsize\color{ftMuted}environmental and sensor records}};

\node[box, text width=4.7cm] (id) at (5.5,6.45)
  {\textbf{Identity layer}\\{\scriptsize\color{ftMuted}persistent site and tree identifiers link measured and projected states over time}};
\node[box, text width=2.1cm, minimum height=1.25cm] (meas) at (4.2,4.3)
  {\textbf{Measured observation}\\{\scriptsize\color{ftMuted}traceable to source data}};
\node[box, text width=2.1cm, minimum height=1.25cm] (scen) at (6.8,4.3)
  {\textbf{Scenario projection}\\{\scriptsize\color{ftMuted}conditional on a model}};
\node[box, text width=4.7cm] (api) at (5.5,2.2)
  {\textbf{Shared data interface}\\{\scriptsize\color{ftMuted}preserves state type and temporal lineage}};

\node[box] (ana) at (11.0,6.45)
  {\textbf{Analysis, modelling and GIS}\\{\scriptsize\color{ftMuted}dashboards and scientific workflows}};
\node[box] (vfs) at (11.0,4.3)
  {\textbf{Virtual forest scenes}\\{\scriptsize\color{ftMuted}geometry generated from tree states}};
\node[box, half] (vr) at (10.08,2.2)
  {\textbf{Interactive exploration}\\{\scriptsize\color{ftMuted}Unreal Engine or browser}};
\node[box, half] (syn) at (11.92,2.2)
  {\textbf{Synthetic LiDAR}\\{\scriptsize\color{ftMuted}annotated virtual scans}};

\node[box, fill=white, text width=4.7cm] (gm) at (5.5,-0.75)
  {\textbf{Forest growth and dynamics model}\\{\scriptsize\color{ftMuted}companion-system prototype}};

\draw[parr] (obs.east) -- (id.west);
\draw[parr] (inv.east) -- (meas.west);
\draw[parr] (col.east) -- (api.west);
\draw[arr] (id.south -| meas.north) -- (meas.north);
\draw[arr] (id.south -| scen.north) -- (scen.north);
\draw[arr] (meas.south) -- (meas.south |- api.north);
\draw[arr] (scen.south) -- (scen.south |- api.north);
\draw[parr] (api.east) -- ++(0.35,0) |- (ana.west);
\draw[parr] (api.east) -- ++(0.35,0) |- (vfs.west);
\draw[arr] (vfs.south -| vr.north) -- (vr.north);
\draw[arr] (vfs.south -| syn.north) -- (syn.north);
\draw[arr] ([xshift=-1.2cm]api.south) -- node[lab, left=2pt] {initial\\conditions} ([xshift=-1.2cm]gm.north);
\draw[arr] ([xshift=1.2cm]gm.north) -- node[lab, right=2pt] {scenario\\projections} ([xshift=1.2cm]api.south);

\node[font=\sffamily\scriptsize, text=ftMuted, anchor=north] at (5.5,-1.65)
  {\textcolor{ftLine}{\rule[0.12em]{0.5cm}{0.9pt}}\; current foundation or companion-system prototype
   \qquad
   \textcolor{ftLine}{\rule[0.12em]{0.14cm}{0.8pt}\,\rule[0.12em]{0.14cm}{0.8pt}\,%
   \rule[0.12em]{0.14cm}{0.8pt}}\; planned product or connection};
\end{tikzpicture}}
  \caption[Planned integration with the Forest Digital Twin Database]{Planned integration of 3Dtrees.earth observations with the Forest Digital Twin Database. Solid outlines identify existing components or companion-system prototypes, and solid arrows show their existing connections. Dashed outlines and arrows indicate planned products and integration pathways. The database and companion workflows provide existing foundations; their integration with 3Dtrees.earth remains planned. Explicit state labels distinguish observations, scenario projections, and synthetic outputs. Projections and synthetic LiDAR are neither observations nor validated outcomes.}
  \label{fig:digital_twin_data_flow}
\end{figure*}

\end{document}